%% file: main.tex
\documentclass[letterpaper,onecolumn,10pt]{article}
\makeatletter
\def\input@path{{texmf/latex/}}
\makeatother
\usepackage{usenix-2020-09}
\usepackage[letterpaper,margin=1in]{geometry}

\usepackage{tikz}
\usepackage{amsmath}
\usepackage{tikz}
\usepackage{amsmath}
\usepackage[normalem]{ulem}
\usepackage{endnotes,microtype,xspace,fancyvrb,multirow}
\usepackage{xcolor}
\usepackage{xspace}
\usepackage{balance}
\usepackage{pifont}
\usepackage{anyfontsize}
\usepackage{booktabs}
\usepackage{fp}
\usepackage{caption}
\usepackage{listings}
\usepackage{float}
\usepackage{tcolorbox}
\tcbuselibrary{listings,skins}
\usepackage{balance}
\usepackage{textcomp}
\usepackage{stmaryrd}
\usepackage{enumitem}
\usepackage{array}
\usepackage{wrapfig}
\usepackage{makecell}
\usepackage{stfloats}

\usepackage{subcaption}
\usepackage{titlesec}
\usepackage{tabularx}
\usepackage[T1]{fontenc}
\usepackage{bbding}  
\usepackage{url}
\usepackage{colortbl}

\usepackage{layout}
\usepackage{multicol}
\usepackage{csquotes}
\usepackage{tabularx}

\usepackage{booktabs}
\usepackage{multirow}

\usepackage{adjustbox}

\usepackage{xparse}
\usepackage{mathtools} 

\usepackage{amssymb}

\input{sections/design/listing-style}

\newcommand{\harness}{\textsc{OSFoundry}\xspace}
\newcommand{\sys}{\textsc{SpecOS}\xspace}
\newcommand{\oldspec}{\textsc{SysSpec}\xspace}
\newcommand{\spec}{\textsc{SysSpec}\textsuperscript{*}\xspace}

\newcommand{\myparagraph}[1]{\noindent\textbf{#1}}

\definecolor{diaozi}{RGB}{93,49,49}
\definecolor{hightcode}{rgb}{1.0,0.13,0.32}
\definecolor{hightcomment}{RGB}{205, 92, 92}
\definecolor{myRubineRed}{RGB}{209, 0, 86}

\definecolor{diaozi}{RGB}{93,49,49}
\definecolor{hightcode}{rgb}{1.0,0.13,0.32}
\definecolor{myRubineRed}{RGB}{209, 0, 86}
\microtypecontext{spacing=nonfrench}

\begin{document}

\date{}


\title{\harness: Building and Evolving Operating Systems with Specification-Guided Agents}

\author{\rm Hengbin Zhang\textsuperscript{1},\; Qingyuan Liu\textsuperscript{1},\; Mo Zou$$\textsuperscript{1},\; Dong Du\textsuperscript{1,2}\textsuperscript{\Envelope},\; Yubin Xia\textsuperscript{1,2},\; Haibo Chen\textsuperscript{1}\\
{\normalsize \textsuperscript{1}\;Institute of Parallel and Distributed Systems} \\
{\normalsize \textsuperscript{2}\;Advanced Research Center for Agent-Oriented OS} \\
{\normalsize Shanghai Jiao Tong University}
}

\maketitle

\begingroup
\renewcommand{\thefootnote}{}%
\footnotetext{\Envelope~Corresponding author: \url{dd_nirvana@sjtu.edu.cn}.}
\endgroup


\input{sections/abstract}

\input{sections/introduction}
\input{sections/motivation}
\input{sections/overview}
\input{sections/design/design}
\input{sections/evaluation/evaluation}

\input{sections/related-work}
\input{sections/conclusion}








\bibliographystyle{plain}

\bibliography{ref}

\end{document}

%% file: sections/design/listing-style.tex
\definecolor{specblockaccent}{RGB}{46, 101, 105}
\definecolor{specblockborder}{RGB}{177, 194, 195}
\definecolor{specblocktitle}{RGB}{232, 240, 240}
\definecolor{specblockbackground}{RGB}{248, 250, 250}
\definecolor{specblocktitletext}{RGB}{35, 55, 57}

\lstdefinelanguage{text}{}

\tcbset{
  specblockbase/.style n args={1}{
    enhanced,
    listing only,
    listing engine=listings,
    colback=specblockbackground,
    colframe=specblockborder,
    colbacktitle=specblocktitle,
    coltitle=specblocktitletext,
    title={#1},
    fonttitle=\sffamily\bfseries\fontsize{8pt}{9.4pt}\selectfont,
    boxrule=0.45pt,
    arc=1mm,
    outer arc=1mm,
    left=2.2mm,
    right=1.8mm,
    top=1.3mm,
    bottom=1.3mm,
    boxsep=0pt,
    titlerule=0.35pt,
    toptitle=1.1mm,
    bottomtitle=1.1mm,
    width=0.96\linewidth,
    center,
    before skip=3pt,
    after skip=0pt,
  },
}

\newtcblisting{specblock}[3][]{
  specblockbase={#2},
  listing options={
    language=#3,
    basicstyle=\ttfamily\fontsize{8.2pt}{9.8pt}\selectfont,
    columns=fullflexible,
    keepspaces=true,
    showstringspaces=false,
    breaklines=true,
    breakatwhitespace=false,
    breakindent=1.5em,
    tabsize=2,
    escapeinside={|}{|}
  },
  #1
}

\newtcblisting{implementationblock}[1][]{
  specblockbase={IMPLEMENTATION},
  listing options={
    language=C,
    basicstyle=\ttfamily\fontsize{8pt}{9.5pt}\selectfont,
    columns=fullflexible,
    keepspaces=true,
    showstringspaces=false,
    numbers=left,
    numbersep=6pt,
    breaklines=true,
    breakatwhitespace=false,
    breakindent=1.5em,
    tabsize=2
  },
  #1
}

%% file: sections/abstract.tex
\begin{abstract}

Operating systems must evolve continuously.
Yet their development remains code-centric and largely manual: even a localized change can require recovering implicit assumptions, coordinating multiple subsystems, and repeatedly building, booting, testing, and debugging the complete system.
General-purpose coding agents automate individual edits, but their prompt-centric workflows repeatedly reconstruct task boundaries and OS semantics from scattered context, limiting their reliability for sustained OS evolution.

This paper presents \harness{}, an OS-specialized agent harness that makes design intent persistent throughout OS development.
Its key insight is to \emph{separate stable intent from diverse implementation}.
Instead of relying on free-form natural-language prompts to convey design intent, \harness{} uses \spec{}, a shared development blueprint comprising a task-bounding Plan and an OS-specific Specification: the Plan bounds what a change should achieve, while the Specification records the interfaces, modular dependencies, and concurrency semantics that its implementation must preserve.
Agents implement and validate code against this same blueprint, refining both \spec{} and the implementation when execution or review exposes a mismatch.

We evaluate \harness{} along three ways.
First, \harness{} generates \sys{} from \spec{}; the resulting complete OS boots and passes all 70 functional tests.
Second, specification patches evolve \sys{} with a GUI and three performance optimizations that improve performance by up to 4.41$\times$.
Third, across 11 recent Linux-kernel bug-fix tasks, \harness{} achieves 1.8$\times$ the accuracy of Codex using GPT-5.5.
These results show that persistent specifications can shift OS construction and evolution from repeated manual kernel engineering toward specification-guided development.

\end{abstract}

%% file: sections/introduction.tex
\section{Introduction}
\label{s:intro}

OS development is increasingly defined by customization and continuous evolution.
Operating systems must provide stable abstractions while adapting to new processor architectures such as RISC-V~\cite{waterman2014risc}, emerging memory interconnects such as CXL~\cite{10.1145/3575693.3578835} and UB~\cite{liao2025ubmeshhierarchicallylocalizedndfullmesh}, specialized devices such as NPUs and LPUs~\cite{nv-lpu}, and changing security and workload requirements.
No single implementation serves all deployment contexts optimally, so developers must continually construct specialized OS variants and evolve mature kernels such as Linux without breaking existing applications.

Yet the dominant OS development model remains code-centric and largely manual.
Developers translate high-level requirements into low-level patches, recover undocumented assumptions from a large codebase, coordinate changes across subsystems, and repeatedly build, boot, test, and debug the result.
This process is difficult not only because kernels are large, but also because OS changes are rarely local: a modification to memory allocation, for example, can affect virtual memory, process creation, page-fault handling, and subsystem-specific performance assumptions.
Consequently, OS development must simultaneously preserve compatibility, respect cross-subsystem contracts, and accommodate deployment-specific design choices.

LLM-based coding agents offer a promising way to reduce this burden, but their prompt-centric interface is poorly matched to systems software.
General-purpose agents such as Codex, Claude Code, and SWE-agent~\cite{codex,claudecode,yang2024swe} can navigate repositories, modify code, and invoke development tools.
However, they repeatedly infer task boundaries and system semantics from a prompt and the surrounding code.
Natural-language descriptions rarely capture precise state transitions, interface assumptions, locking protocols, or hardware-facing invariants, while repository context scatters this information across implementations, tests, documentation, and comments.
As a result, locally plausible code may compile yet violate a remote dependency or fail only when the complete OS boots and executes.

\begin{figure}[t]
  \setlength{\belowcaptionskip}{-8pt}
  \setlength{\abovecaptionskip}{0pt}
  \centering
  \includegraphics[width=0.98\textwidth]{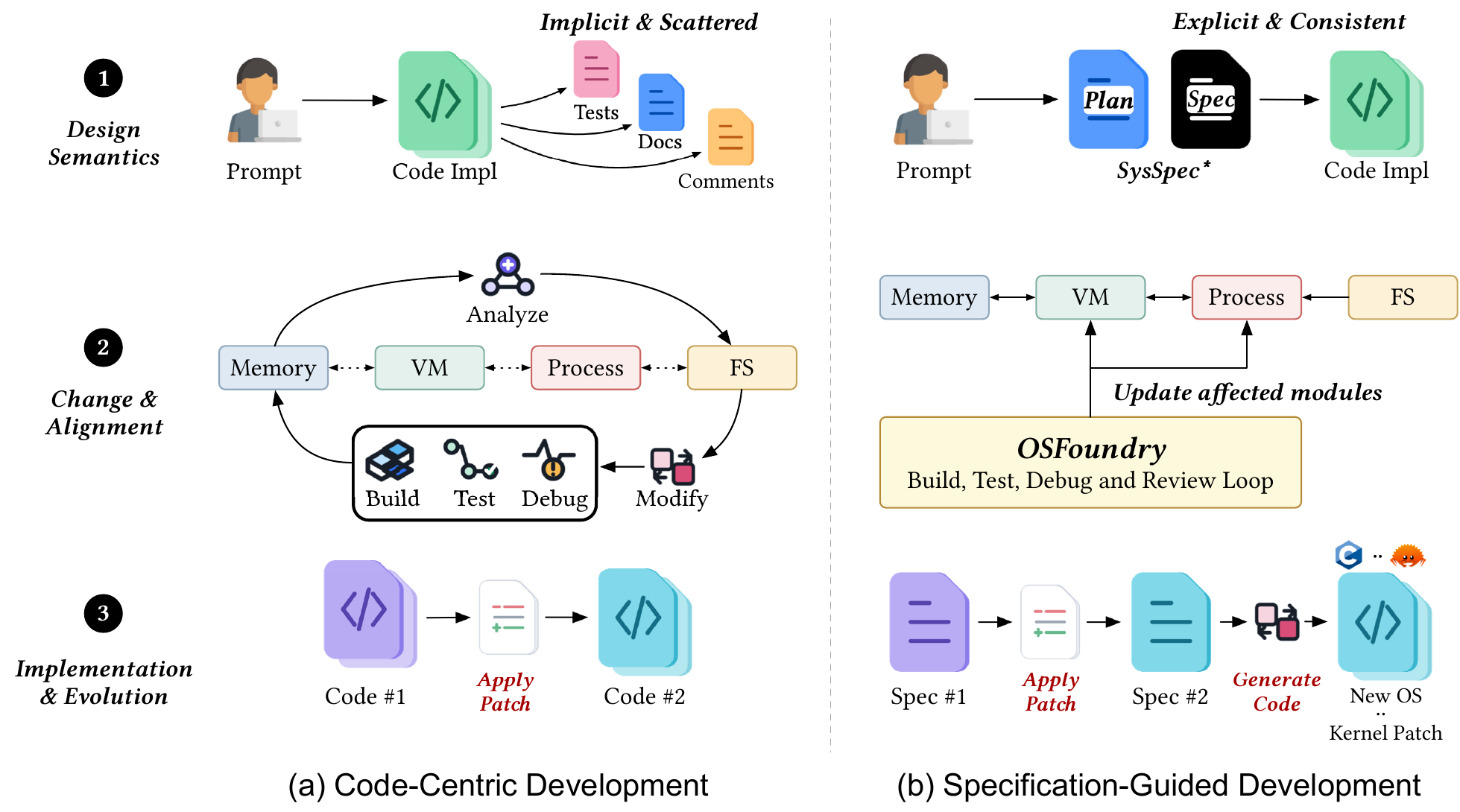}
  \caption{\textbf{Code-centric OS development versus \harness{}.}
  (a) Code-centric development scatters design intent across prompts, source code, tests, documentation, and comments, leaving developers to coordinate changes across the OS stack.
  (b) \harness{} captures developer intent in \spec{} that records OS semantics, then uses this shared development blueprint to coordinate implementation, validation, review, and refinement.}
  \label{fig:design-comparison}
\end{figure}

SpecFS~\cite{315951} established that structured specifications can provide a more reliable interface between developer intent and LLM-generated systems code.
Its \oldspec{} framework captured functionality, modularity, and concurrency semantics and used them to generate and evolve a complete file system.
Nevertheless, SpecFS stopped at a single user-space subsystem: it did not exercise kernel-mode execution, privilege boundaries, hardware interfaces, or contracts spanning the full OS stack.
Its workflow also began with developer-authored specifications and therefore did not address how agents should extract task-relevant semantics from an existing kernel or preserve the resulting blueprint throughout implementation, review, and evolution.

This paper presents \harness{}, an OS-specialized agent harness for both constructing new operating systems and developing or optimizing existing kernels.
Its key insight is to \emph{separate stable design intent from replaceable implementation} by making design intent a first-class artifact shared by every development stage.
As illustrated in \autoref{fig:design-comparison}, \harness{} does not ask each agent to reconstruct OS semantics independently from a free-form prompt.
Instead, it uses \spec{} as a persistent development blueprint shared across development stages.
Every stage uses this blueprint to guide its decisions, while \harness{} refines both the blueprint and code as the task evolves.

\spec{} comprises a task-bounding \emph{Plan} and an OS-specific \emph{Specification}.
The Plan states the goals and non-goals, permissible codebase scope, and observable success criteria for a task.
The Specification records the semantics that an implementation must satisfy: functionality specifications describe state transitions and invariants, modularity specifications define imported assumptions and exported guarantees, and concurrency specifications make synchronization and ordering constraints explicit.
For a new OS, this blueprint can govern the complete system stack; for an existing kernel, the specification agent extracts and refines only the semantics relevant to the intended change.
The resulting blueprint is precise and inspectable, but does not claim a machine-checked proof.

Four specialized agents use this shared blueprint in a closed development loop.
The specification agent constructs \spec{} from developer input and code analysis by producing its Plan and Specification; the implementation agent realizes the Specification as code under the Plan's constraints and exercises it through build and test workflows; the review agent performs general and OS-specific semantic review; and the refinement agent incorporates review findings and developer feedback before the next iteration.
A shared OS knowledge base supplies domain knowledge, while an executable environment with build tools, QEMU, and GDB grounds decisions in compilation, boot, test, and debugging evidence.

\harness{} maintains full-system semantic context without requiring every agent to process the complete codebase at once.
It organizes the Specification within \spec{} into dependency-aware, context-bounded SpecModules, each of which contains enough functional and concurrency context for local implementation while exposing explicit rely-guarantee contracts to other modules.
When a requirement changes, a specification patch updates the affected modules and propagates revised guarantees through their dependents while leaving unrelated contracts intact.
This mechanism enables both generative operating systems, whose implementations are derived from a complete specification, and scoped evolution of existing kernels, where only the task-relevant blueprint and code must change.

We demonstrate full-system construction and evolution with \sys{}, a complete operating system described by approximately 8K lines of \spec{} and organized into specification-governed modules.
The generated system boots, runs a standard user environment, and passes its test suite.
We then evolve the text-based system into a GUI-capable OS through specification-level patches spanning kernel graphics support, user-space services, libraries, and applications.
We further apply performance-oriented patches for a per-CPU page allocator, copy-on-write fork, and batched pipe I/O; the resulting implementations preserve their external contracts while improving performance by up to 4.41$\times$.

We separately evaluate whether the same architecture improves development in an existing kernel.
Across 11 representative Linux-kernel bug-fix tasks, \harness{} achieves 1.8$\times$ the accuracy of Codex using GPT-5.5.
Together, the two evaluations show that specification-guided agents can move beyond isolated code generation: they can construct and evolve a complete OS while also making scoped development in a mature kernel more accurate and reliable.

This paper makes the following contributions:

\begin{itemize}[leftmargin=*]

\item We introduce \harness{}, an OS-specialized agent harness that makes developer intent a persistent, first-class artifact shared across specification, implementation, review, and refinement. Its \spec{} development blueprint comprises a task-bounding Plan and an OS-specific Specification, while a shared knowledge base and executable environment support both complete OS construction and scoped development of existing kernels.

\item We extend the specification-guided approach of SpecFS from a single subsystem to the full OS stack and implement \sys{}, a complete OS described by approximately 8K lines of \spec{}. \sys{} boots, runs a standard user environment, and evolves through specification patches to support a GUI and multiple performance optimizations, including one that improves performance by up to 4.41$\times$.

\item We evaluate \harness{} on 11 representative Linux-kernel bug-fix tasks. Compared with Codex using GPT-5.5, \harness{} achieves 1.8$\times$ the accuracy.

\end{itemize}

%% file: sections/motivation.tex
\section{The Case for Agentic OS Development}

\subsection{Growing Demand for OS Customization}

OS customization is a fundamental requirement rather than a niche optimization.
Computing platforms span embedded devices with kilobytes of memory, mobile systems governed by energy and security constraints, and warehouse-scale services operating across millions of cores.
No single OS configuration or implementation can serve these environments optimally.
Each target instead requires a particular composition of scheduling policies, memory-management strategies, I/O paths, device support, and security mechanisms.

Today, deriving these purpose-built variants remains largely manual.
Developers typically fork or configure an existing kernel, add platform-specific modules, and patch subsystem implementations until the resulting system satisfies a target workload.
Because the intended design is distributed across configuration files, source code, tests, and documentation, reproducing or transferring a customization to another variant requires substantial engineering effort.

\begin{figure*}[htb]
  \setlength{\belowcaptionskip}{-8pt}
  \setlength{\abovecaptionskip}{0pt}
  \centering
  \begin{minipage}[b]{0.53\textwidth}
    \centering
    \includegraphics[width=0.96\linewidth]{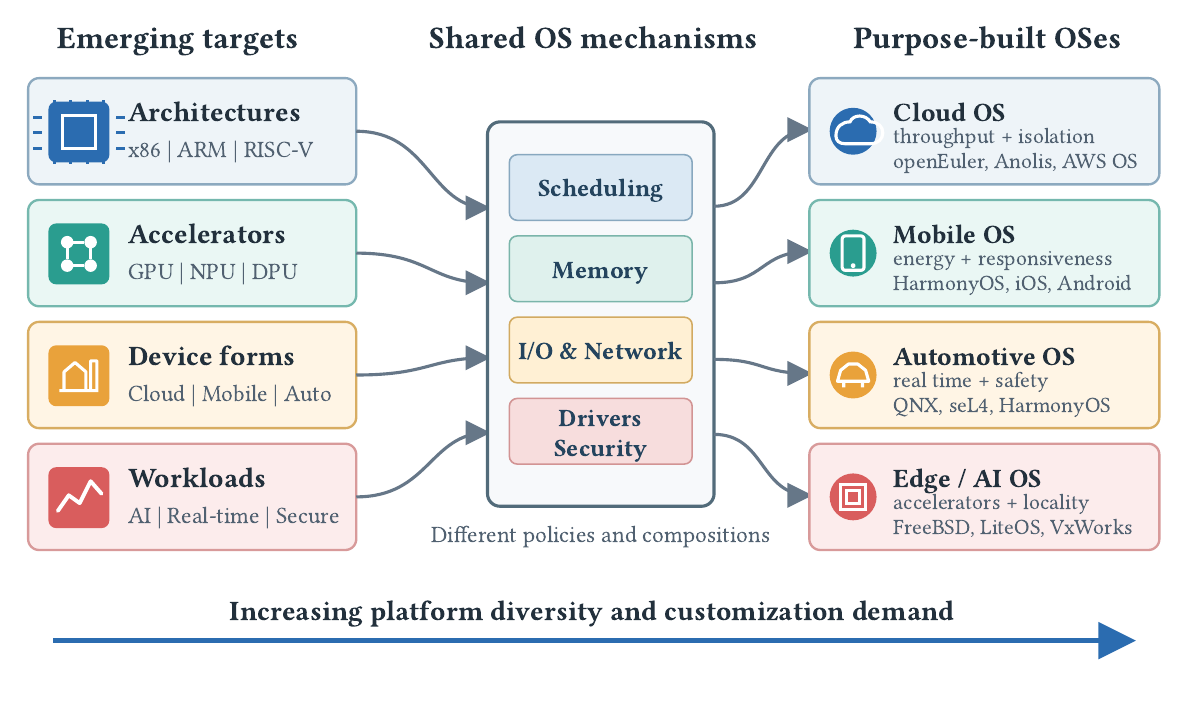}
    \smallskip
    {\small \textbf{(a)} The need for OS customization.}
  \end{minipage}
  \hfill
  \begin{minipage}[b]{0.43\textwidth}
    \centering
    \includegraphics[width=0.96\linewidth]{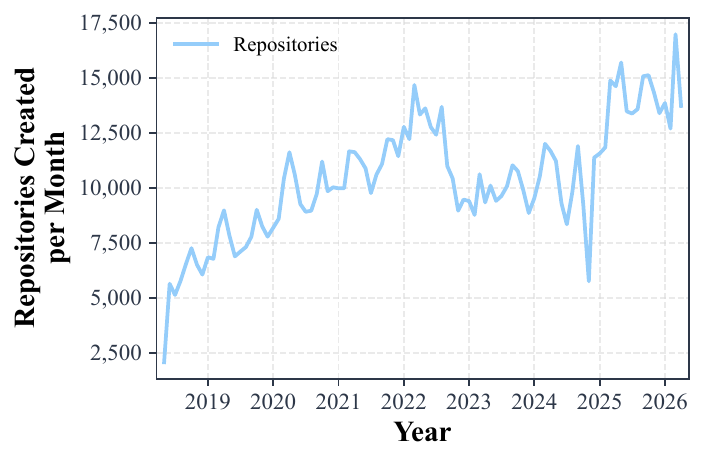}
    \smallskip
    {\small \textbf{(b)} New OS-related repositories per month.}
  \end{minipage}
  \caption{\textbf{Growing demand for OS customization.}
  \textit{(a) Diverse architectures, accelerators, device forms, and workloads impose different requirements on shared OS mechanisms, motivating purpose-built variants.
  (b) The sustained growth in newly created OS-related GitHub repositories provides empirical evidence of this demand.}}
  \label{fig:motiv-customization}
\end{figure*}

The demand for such customization is both broad and growing.
As summarized in \autoref{fig:motiv-customization}(a), emerging architectures, accelerators, device forms, and workloads all impose different policies on a common set of OS mechanisms.
Our study of OS-related GitHub repositories over the past eight years further shows sustained growth in newly created repositories (\autoref{fig:motiv-customization}(b)).
The broader ecosystem of Linux distributions and vendor-maintained Android variants reinforces the same point: developers repeatedly assemble different systems from closely related subsystem foundations.

This diversity creates a compounding cost.
A purpose-built OS is not finished once it boots; it must continue to absorb new hardware support, security defenses, optimizations, and application requirements.
The practical challenge is therefore not merely producing more variants, but evolving many variants without repeatedly rediscovering and reimplementing the same system semantics.

\subsection{Why OS Evolution Is Difficult}
\label{ssec:os-evolution}

\input{sections/tables/os-evolution-features}

\subsubsection{Forces Driving Kernel Evolution}

OS evolution is continuous because the environment beneath and above the kernel never stops changing.
Linux has evolved for more than three decades, growing from roughly 10,000 lines in version 0.01 (1991) to tens of millions of lines today~\cite{linux-src}.
\autoref{tab:os-evolution-features} illustrates this process through representative features organized by six recurring technological forces.

\myparagraph{Hardware scale and data movement.}
The transition from single-core to many-core processors, from uniform memory to NUMA and persistent memory, and from disks to NVMe has repeatedly forced redesigns of scheduling, memory management, and storage paths.
Network line rates have similarly driven multi-core packet steering, in-driver processing, and new congestion-control mechanisms.
These changes alter not only individual algorithms, but also the concurrency and data-placement assumptions shared across subsystems.

\myparagraph{Deployment and policy diversity.}
Virtualized and containerized clouds require strong isolation, resource accounting, and device virtualization, while mobile systems make energy management and low-latency IPC first-class concerns.
The same kernel must therefore support substantially different resource policies and trusted computing boundaries.
This pressure expands both the number of mechanisms and the configurations in which they must coexist.

\myparagraph{Security and programmability.}
New attacks continually add defense layers such as mandatory access control, syscall filtering, address randomization, and microarchitectural mitigations.
At the same time, eBPF and related mechanisms expose controlled programmability inside performance-critical kernel paths.
Both trends increase the semantic surface that future changes must preserve.

These forces are persistent rather than episodic.
As hardware and workloads continue to diversify, OS developers must repeatedly introduce new mechanisms while retaining the behavior on which existing applications and deployments depend.

\subsubsection{Three Structural Challenges}

\begin{figure*}[htb]
  \setlength{\belowcaptionskip}{-8pt}
  \setlength{\abovecaptionskip}{0pt}
  \centering
  \includegraphics[width=0.98\textwidth]{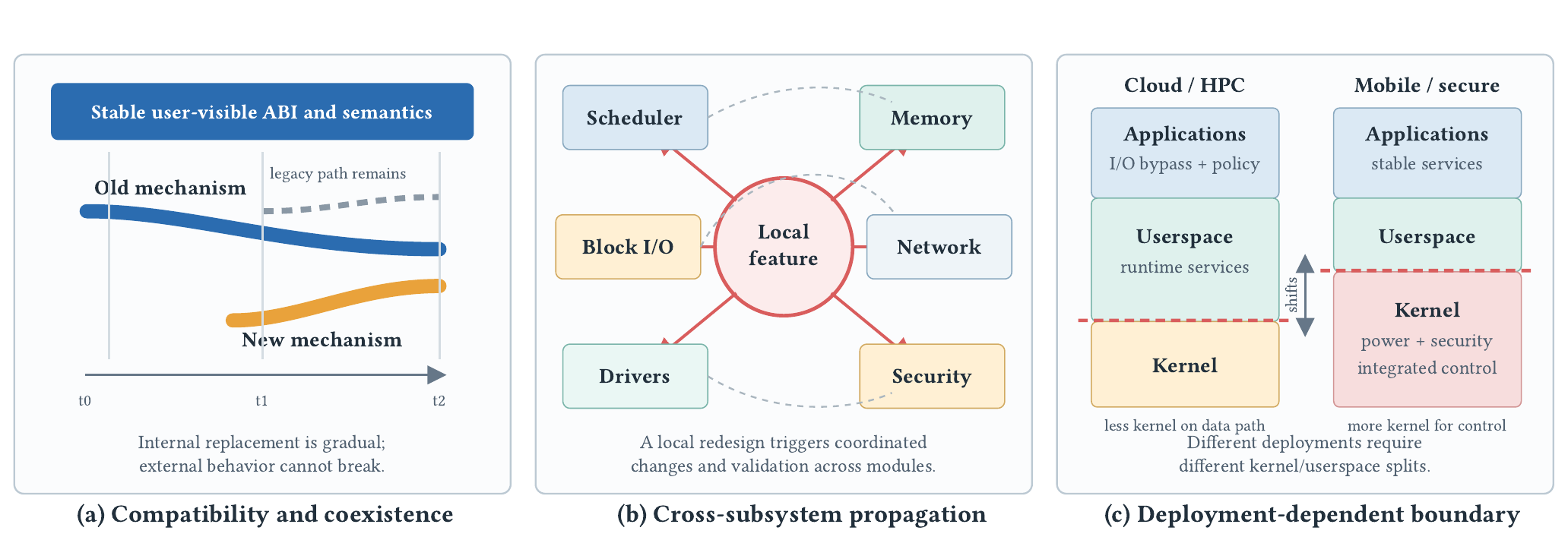}
  \caption{\textbf{Why OS evolution is difficult.}
  \textit{(a) New mechanisms must coexist with legacy paths while preserving user-visible semantics.
  (b) A local feature can trigger coordinated changes across tightly coupled subsystems.
  (c) Different deployments place functionality on different sides of the kernel/userspace boundary, multiplying the configurations and contracts that must be maintained.}}
  \label{fig:motiv-evolution-challenges}
\end{figure*}

The difficulty of OS evolution lies not only in \emph{what} changes, but in three recurring properties of \emph{how} change must occur, summarized in \autoref{fig:motiv-evolution-challenges}.

\myparagraph{Compatibility turns replacement into coexistence.}
New kernel mechanisms rarely replace old ones abruptly.
The kernel must preserve user-visible interfaces and behavioral semantics, so migration proceeds incrementally and old and new paths may coexist for years.
Scheduler evolution illustrates this constraint: CFS replaced the O(1) scheduler while preserving the \texttt{sched\_class} abstraction, and EEVDF later changed the core scheduling algorithm while retaining the external class hierarchy and cgroup integration.
Internal implementation can change substantially, but externally visible guarantees cannot.

\myparagraph{Subsystem coupling makes local changes non-local.}
Although a kernel is organized into named subsystems, significant features routinely cross their boundaries.
The cgroups v2 redesign, for example, required changes not only to the cgroup core but also to resource controllers, the page and slab allocators, the OOM killer, scheduling, block I/O, and networking.
The pace and correctness of one change therefore depend on maintainers identifying every affected contract and coordinating updates across modules.

\myparagraph{Deployment diversity shifts the kernel/userspace boundary.}
Different workloads place conflicting functionality on either side of the privilege boundary.
DAX and io\_uring move data-path work away from conventional kernel mediation, whereas eBPF, XDP, and sched\_ext place user-defined policies inside kernel execution paths.
Cloud deployments may favor a smaller kernel role on throughput-critical paths, while mobile systems depend on integrated kernel control for energy management and security.
Supporting these choices multiplies code paths, interactions, and semantic contracts.

Together, compatibility, subsystem coupling, and deployment diversity make OS evolution a semantic-coordination problem.
Writing a locally plausible function is insufficient; developers must preserve contracts across versions, modules, privilege levels, and deployment configurations.
In a code-centric workflow, these contracts remain implicit and scattered, which is precisely the information an agentic development system must make explicit.

\subsection{Why Prior Efforts Fall Short}
\label{ssec:prior-efforts}

\input{sections/tables/related-work}

Prior work addresses important pieces of agentic software development, specification-driven engineering, and verified OS construction, as summarized in \autoref{tab:conceptual-comparison}.
However, no prior approach in the table connects task intent, OS semantics, implementation, and execution evidence through one shared development blueprint.
The missing capability is therefore architectural rather than simply a matter of using a stronger language model.

\myparagraph{General-purpose harnesses coordinate actions, but not OS semantics.}
Coding agents such as Codex, Claude Code, and SWE-agent can navigate repositories, modify code, and invoke development tools, while AHE and VeRO optimize the agent scaffold itself.
These systems improve how an agent operates, but do not define the syscall behavior, interface assumptions, locking protocols, or hardware invariants that a particular OS task must preserve.
Each implementation or review stage may consequently reconstruct a different interpretation of the same change.

\myparagraph{Workflow specifications are too generic, while verified OSes require proof engineering.}
Spec Kit and Kiro persist requirements, plans, and tasks, but their artifacts do not natively encode OS functionality, modular dependencies, and concurrency semantics.
At the other end of the design space, seL4, CertiKOS, and Hyperkernel provide stronger machine-checked guarantees at the cost of formal models and substantial proof or specification engineering.
An agentic OS workflow needs a practical middle ground: artifacts precise enough to guide implementation and semantic review, but lightweight enough to author or extract during ordinary development.

\myparagraph{OS-oriented coding agents remain task-specific.}
Existing efforts use LLMs for kernel patch generation, fault localization, or specification synthesis.
These are valuable capabilities, but their outputs do not form a shared development blueprint that persists from task definition through implementation, whole-system validation, review, and later refinement.
Nor do they provide one workflow that supports both constructing a complete OS and modifying an existing kernel.

These comparisons expose four requirements for end-to-end agentic OS development.
The system must persist task boundaries and OS semantics, enable local reasoning without losing cross-module dependencies, ground decisions in build and execution evidence, and refine the blueprint as requirements or review findings evolve.
The next subsection examines why SpecFS provides a promising semantic foundation and which gaps remain before these requirements can be met.


\subsection{\oldspec{}: Foundation and Limitations}
\label{sec:specfs-to-specos}

SpecFS~\cite{315951} established that structured specifications can serve as a precise interface between system designers and LLM-based coding agents.
It introduced \oldspec{}, a framework for specification-guided system construction and evolution, and instantiated the framework in a concurrent file system.
Rather than asking an LLM to recover system semantics from a free-form prompt or an entire codebase, developers encode design intent in structured, formal-method-inspired contracts that guide generation, review, and evolution.
\oldspec{} is deliberately not a proof language; it uses structured natural language and type annotations while adapting disciplined concepts from Hoare logic and rely-guarantee reasoning.

\myparagraph{Structured semantic contracts.}
\oldspec{} models a system as context-bounded modules governed by three complementary specifications.
A \emph{functionality specification} describes state transitions through Hoare-style preconditions, postconditions, invariants, and, when necessary, a high-level algorithm~\cite{zou2019atomfs,chen2015using}.
A \emph{modularity specification} makes dependencies explicit through adapted rely-guarantee contracts~\cite{10.1145/1480881.1480922}: Rely clauses state imported types, functions, and assumptions, while Guarantee clauses define exported interfaces and behavior.
A separate \emph{concurrency specification} records lock ownership, ordering constraints, and concurrency-related preconditions and postconditions.
Together, these contracts provide enough local context to implement one module while retaining the semantics required for composition.

\myparagraph{Specification-guided construction and evolution.}
\oldspec{} treats the specification, rather than generated C code, as the primary design artifact.
A DAG-structured \emph{spec patch} decomposes a feature into localized changes and makes propagation dependencies explicit: leaf nodes introduce self-contained changes, intermediate nodes update dependent guarantees, and root nodes preserve externally visible behavior.
The SpecFS toolchain couples specification authoring, module-level generation, specification-aware validation, and retry with actionable feedback.
This loop mitigates LLM failures without assuming that generated code is intrinsically reliable.

\myparagraph{What SpecFS demonstrated.}
SpecFS described the high-level design of AtomFS~\cite{zou2019atomfs} in 45 specification-governed modules and generated approximately 4.3K lines of C.
Its user-space FUSE implementation matched AtomFS's test profile across 754 \texttt{xfstests} cases, with remaining failures corresponding to functionality outside the prototype's scope.
SpecFS also evolved the design through ten representative Ext4 features using specification patches.
These results established that structured semantic contracts can guide LLMs in constructing and evolving a complete OS subsystem.

\myparagraph{What SpecFS left open.}
SpecFS nevertheless stopped at a single user-space subsystem.
It did not exercise kernel-mode execution, hardware-facing interfaces, privilege transitions, or contracts spanning memory management, processes, system calls, device drivers, and the user runtime.
Its workflow began with developer-authored draft specifications, leaving specification extraction from an existing codebase unaddressed.
Finally, file-system-level validation could not expose build, boot, or full-system integration failures that emerge only when kernel and user modules execute together.
SpecFS therefore did not establish whether specification-guided agents can scale to a complete OS or improve development in an existing kernel.

\subsection{Opportunity: Consistent Design Intent for Agentic OS Development}
\label{ssec:opportunity}

The preceding evidence points to a broader opportunity.
OS customization creates sustained demand for specialized systems, OS evolution requires coordination across implicit semantic contracts, existing agentic tools cover only fragments of the lifecycle, and SpecFS shows that structured specifications can make system semantics actionable for LLMs.
The missing abstraction is a persistent representation of design intent shared across every stage of the OS development lifecycle.

The key insight is to \textbf{\emph{separate stable design intent from replaceable implementation}}.
\spec{}, an extension of \oldspec{}, forms this shared development blueprint.
It comprises a task-bounding Plan, which records goals, non-goals, codebase scope, and success criteria, and an OS-specific Specification, which records the functionality, modularity, and concurrency semantics that an implementation must preserve.
Implementation, execution, review, and refinement can then operate on the same interpretation of developer intent rather than reconstructing it independently.

This blueprint enables two complementary forms of agentic OS development.
For a new system, a complete collection of modular specifications defines a \emph{generative OS} (\sys in \autoref{subs:design-put-it-together}): a family of implementations that can be synthesized, customized for a target, and evolved by updating specifications while preserving external guarantees.
For an existing kernel (e.g., Linux kernel), the blueprint can instead be task-bounded: agents extract and refine only the semantics relevant to a feature or optimization, then use build, boot, test, and review evidence to validate the resulting patch.
Existing-kernel development therefore does not require rewriting or specifying the entire system.

This reasoning leads directly to \harness{}, an OS-specialized agent harness that keeps \spec{}, implementation, and validation evidence aligned throughout the development lifecycle.
Its purpose is not to replace OS developers with LLMs, but to elevate their role by changing how design intent is expressed, preserved, and evolved.

%% file: sections/tables/os-evolution-features.tex
\begin{table*}[!htb]
  \centering
  \caption{Representative Linux kernel features and their evolutionary drivers.}
  \label{tab:os-evolution-features}
  \small
  \begin{tabular}{@{}p{0.18\textwidth} p{0.22\textwidth} p{0.13\textwidth} p{0.37\textwidth}@{}}
    \toprule
    \textbf{Motivation} & \textbf{Feature} & \textbf{Kernel Version} & \textbf{Description} \\
    \midrule
    Multicore \& Scalability
      & O(1) Scheduler $\rightarrow$ CFS & 2.6.0 / 2.6.23 & Fair scheduling for multi-core systems \\
      & Transparent Huge Pages & 2.6.38 & Automatic huge pages to reduce TLB misses \\
      & AutoNUMA & 3.8 & Automatic page migration for NUMA locality \\
      & blk-mq & 3.13 & Multi-queue block layer for parallel I/O \\
      & DAX & 4.0 & Direct access to persistent memory, bypassing page cache \\
      & io\_uring & 5.1 & Async I/O via shared ring buffers \\
      & EEVDF Scheduler & 6.6 & Deadline-fair scheduling replacing CFS \\
    \midrule
    Virtualization \& Cloud Computing
      & KVM & 2.6.20 & In-kernel hypervisor for hardware virtualization \\
      & PID \& Net Namespaces & 2.6.24 & Process and network isolation for containers \\
      & Cgroups v1 / v2 & 2.6.24 / 4.5 & Hierarchical resource control; v2 unified hierarchy \\
      & OverlayFS & 3.18 & Layered filesystem for container images \\
      & VFIO & 3.6 & IOMMU-backed direct device assignment to VMs \\
    \midrule
    Network Bandwidth \& Latency
      & RPS / RFS / XPS & 2.6.35 & Multi-core packet steering in software \\
      & nftables & 3.13 & Flexible successor to iptables \\
      & XDP & 4.8 & BPF packet processing at driver level \\
      & BBR & 4.9 & Model-based congestion control \\
      & MPTCP & 5.6 & Multi-path TCP for bandwidth aggregation \\
    \midrule
    Mobile \& Power Efficiency
      & Tickless Kernel & 2.6.21 & Eliminate idle timer interrupts \\
      & Wakelocks & 3.5 & Suspend-aware power management for mobile \\
      & Binder IPC & 3.19 & Android's high-performance IPC mechanism \\
      & schedutil Governor & 4.7 & Scheduler-driven CPU frequency scaling \\
      & PSI & 4.20 & Unified resource pressure metrics \\
    \midrule
    Security Threats
      & SELinux & 2.6.0 & Mandatory access control \\
      & seccomp & 2.6.12 & System call filtering \\
      & KASLR & 3.14 & Kernel address randomization \\
      & KPTI & 4.15 & Kernel/user page table isolation (Meltdown fix) \\
      & Retpoline & 4.15 & Spectre v2 indirect branch mitigation \\
      & Landlock & 5.13 & Unprivileged stackable sandboxing \\
    \midrule
    Programmability
      & eBPF Verifier + Helpers & 3.18--5.x & Safe in-kernel VM for user-defined programs \\
      & BPF CO-RE & 5.3 & Portable BPF across kernel versions \\
      & sched\_ext & 6.12 & BPF-programmable scheduling policies \\
    \bottomrule
  \end{tabular}
\end{table*}

%% file: sections/tables/related-work.tex
\begin{table*}[!htb]
  \centering
  \small
  \setlength{\tabcolsep}{5pt}
  \renewcommand{\arraystretch}{1.12}
  \caption{Review of prior work and comparison with the design of \harness{}.}
  \label{tab:conceptual-comparison}
  \begin{tabularx}{\textwidth}{
    >{\raggedright\arraybackslash}p{0.14\textwidth}
    >{\hsize=1.2\hsize\linewidth=\hsize\raggedright\arraybackslash}X
    >{\hsize=0.8\hsize\linewidth=\hsize\raggedright\arraybackslash}X}
    \toprule
    \textbf{} & \textbf{Prior Work} & \textbf{Design in \harness{}} \\
    \midrule

    \textbf{Agent Harness and Workflow}
    & \underline{\textit{Codex}}~\cite{codex},
      \underline{\textit{Claude Code}}~\cite{claudecode}, and
      \underline{\textit{SWE-agent}}~\cite{yang2024swe}
      provide general-purpose capabilities for repository navigation, code modification, and development-tool invocation;
      \underline{\textit{AHE}}~\cite{lin2026agentic} and
      \underline{\textit{VeRO}}~\cite{ursekar2026vero} 
      treat the agent harness itself as an optimization target, adapting its workflow and scaffolding over time.
      \par\textbf{Analysis:} The former lack OS-specific contracts, while the latter
      optimize the agent scaffold rather than target-system semantics.
    & An OS-specialized multi-agent loop keeps specification, implementation,
      review, and refinement aligned around shared system semantics.
      \\
    \midrule

    \textbf{Specification-Driven Development}
    & \underline{\textit{Spec Kit}}~\cite{githubspeckit}, and
      \underline{\textit{Kiro}}~\cite{kiro} persist requirements,
      specifications, plans, and tasks, with repository grounding and validation.
      \par\textbf{Analysis:} Their artifacts are generic and do not natively encode
      OS functional, modular, and concurrency semantics.
    & An OS-specific \spec{} couples a task-bounding Plan with a Specification
      for OS semantics, keeping
      development intent and system semantics aligned with the implementation.
      \\
    \midrule

    \textbf{OS Development Scope}
    & For bug fixing,
      \underline{\textit{kAgent}}~\cite{mathai2026kagent} and
      \underline{\textit{Code Researcher}}~\cite{singh2025code}
      use crash evidence and repository context to generate Linux-kernel patches.
      For fault localization,
      \underline{\textit{LinuxFL+}}~\cite{zhou2026taming}
      uses bug reports and kernel evidence to localize faulty files.
      For specification generation,
      \underline{\textit{OSVBench}}%
      ~\cite{li2025osvbenchbenchmarkingllmsspecification} evaluates LLMs on
      synthesizing complete Hyperkernel state-transition specifications.
      \par\textbf{Analysis:} These approaches target individual kernel-development tasks
      rather than end-to-end OS development.
    & One specification-guided workflow supports both complete OS construction and 
      the evolution of existing kernel code.
      \\
    \midrule

    \textbf{Verification and Validation}
    & \underline{\textit{seL4}}~\cite{klein2009sel4},
      \underline{\textit{CertiKOS}}~\cite{gu2016certikos}, and
      \underline{\textit{Hyperkernel}}~\cite{nelson2017hyperkernel} use
      machine-checked refinement or automated verification.
      \par\textbf{Analysis:} They provide stronger correctness guarantees, at the
      cost of formal models and substantial proof or specification engineering.
    & \harness{} does not provide machine-checked proofs; it uses structured
      semantic contracts, execution feedback, and semantic review for
      lightweight validation.
      \\

    \bottomrule
  \end{tabularx}
\end{table*}

%% file: sections/overview.tex
\section{Design Overview}
\label{s:overview}

\begin{figure*}[htb]
  \centering
  \includegraphics[width=0.95\textwidth]{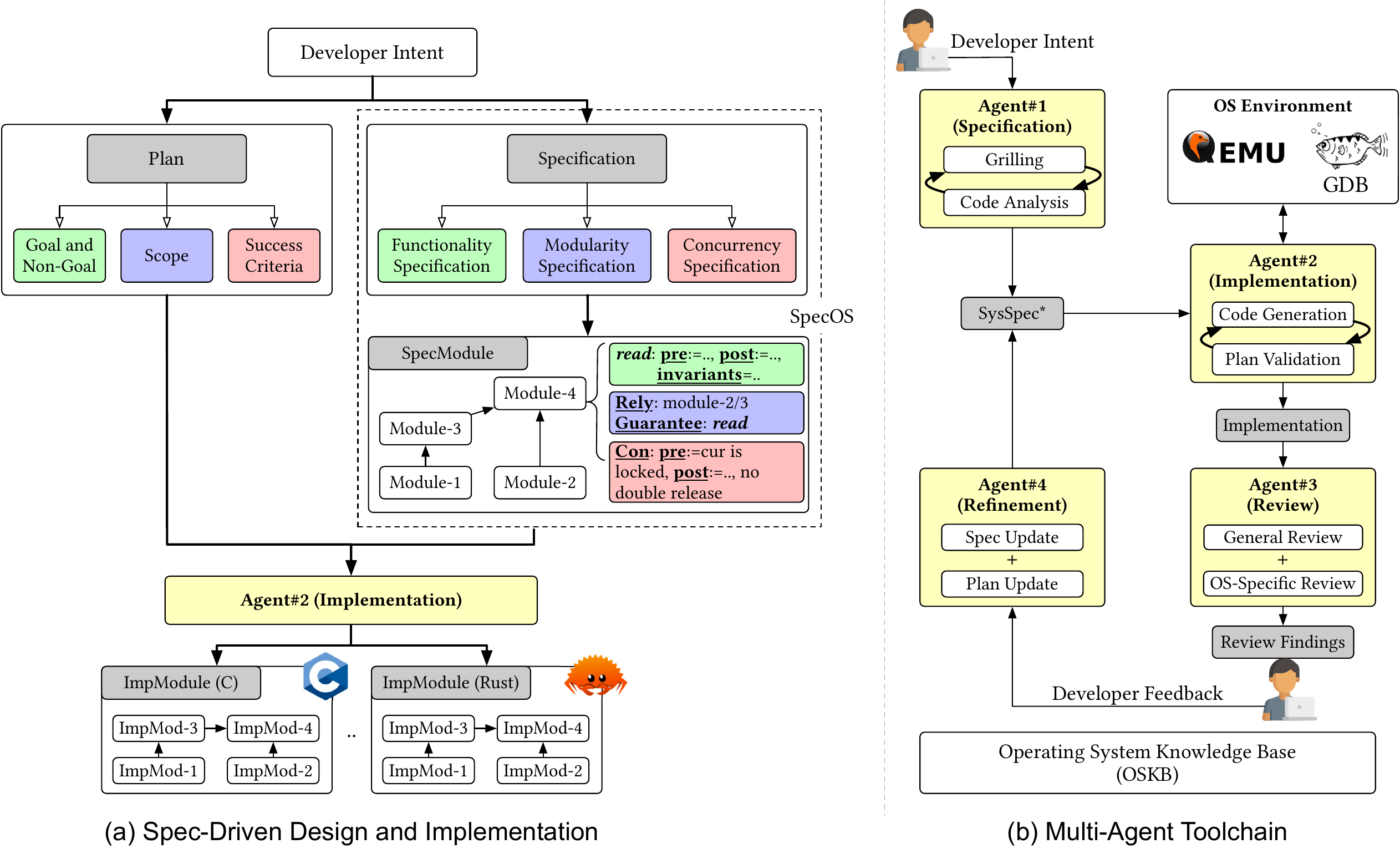}
  \caption{\textbf{Design overview of \harness{}.}
  (a)~Developer intent is captured by \spec{}, which comprises a task-bounding
  \emph{Plan} and an OS-specific \emph{Specification}; the latter's functionality,
  modularity, and concurrency specifications form dependency-aware \emph{SpecModules}; an implementation
  agent realizes these modules in a target systems language.
  (b)~Four specialized agents construct the development blueprint, implement it
  in an executable OS environment, review the resulting code, and refine the
  blueprint using review findings and developer feedback. All agents share an
  operating system knowledge base (OSKB).}
  \label{fig:design-overview}
\end{figure*}

This paper presents \harness{}, an OS-specialized agent harness for both constructing a new operating system and developing or optimizing an existing kernel such as Linux.
As shown in \autoref{fig:design-overview}, \harness{} makes developer intent persistent through a shared development blueprint.
Every stage uses this blueprint to guide its decisions, while \harness{} refines both the blueprint and code when the task evolves.
For a new OS, the blueprint can govern the complete system stack; for an existing kernel, it bounds the intended change and captures only the relevant system semantics.

\myparagraph{Insight: separate stable design intent from replaceable implementation.}
The key insight behind \harness{} is that reliable agentic OS development requires making \emph{design intent} a first-class artifact shared by every stage of development.

General-purpose agents repeatedly infer behavior, change boundaries, and OS invariants from a prompt and a large codebase;
a locally plausible edit can therefore violate a remote interface, locking protocol, or hardware-facing invariant.
\harness{} externalizes this information in a structured, formal-method-inspired blueprint that all agents consume during generation, execution, review, and refinement.
This blueprint is precise and inspectable, but does not claim a machine-checked proof.

\myparagraph{Harness architecture.}
\harness{} comprises three mutually reinforcing layers: an \emph{intent layer} centered on \spec{}, with a task-bounding Plan and an OS-specific Specification as its two components (\autoref{fig:design-overview}(a));
an \emph{agent layer} containing the four specification, implementation, review, and refinement roles; and an \emph{infrastructure layer} that grounds their closed loop in a shared OSKB and an executable environment with QEMU and GDB (\autoref{fig:design-overview}(b)).
Together, these layers address four challenges that arise when moving from a general-purpose coding agent to an end-to-end OS development harness.

\myparagraph{Challenge 1: How can developer intent become an actionable OS spec?}
A natural-language request rarely states non-goals, permissible changes, completion conditions, or implicit OS semantics.
\harness{} addresses this ambiguity through the two components of \spec{}.
The \emph{Plan} defines (1)~goals and non-goals, (2)~codebase scope and allowed dependencies, and (3)~success criteria as observable properties and executable checks.
The \emph{Specification} records the behavior that the implementation must preserve:
functionality specifications define state transitions with preconditions, postconditions, invariants, and, when needed, high-level algorithms;
modularity specifications define imported assumptions and exported guarantees;
and concurrency specifications isolate locking, synchronization, and ordering constraints.

Through iterative grilling and code analysis, the specification agent can construct \spec{} for a new system or extract and refine its task-relevant components around an existing implementation.

\myparagraph{Challenge 2: How can agents reason locally without losing full-system semantics?}
An OS is too large to place in one generation context, yet generating one component in isolation can break dependencies across memory management, processes, file systems, device drivers, and the user runtime.
\harness{} resolves this tension by organizing the Specification as a graph of context-bounded \emph{SpecModules}.
Each module contains the functionality and concurrency semantics needed for local implementation, while its modularity specification names the other modules on which it relies and the guarantees it provides to dependents.
These explicit contracts preserve cross-module context without exposing the entire codebase to every generation step.
The implementation agent can consequently synthesize and revise modules one at a time, compose them according to their dependencies, and target different systems languages, including C and Rust.

\myparagraph{Challenge 3: How can plausible code be turned into an executable OS implementation?}
Even with \spec{}, LLM-generated code may still fail to compile, boot, or satisfy subtle interface and synchronization requirements.
The implementation agent therefore couples code generation with Plan validation and exercises candidates in a realistic OS environment using the available build, test, QEMU, and GDB workflows.
A separate review agent then performs two complementary checks: general code review detects conventional implementation defects, while OS-specific review checks the candidate against the functionality, interface, and concurrency semantics in the Specification.
The shared OSKB supplies consistent domain knowledge and codebase conventions to all agents.
Together, Plan validation, execution feedback, and specification-aware review provide lightweight defense in depth without claiming formal verification.

\myparagraph{Challenge 4: How can the harness support evolution rather than one-shot generation?}
Review findings often expose an underspecified design rather than an isolated coding mistake, while developer feedback may intentionally change the task itself.
The refinement agent therefore updates \spec{} by refining its Plan and affected parts of its Specification: it refines task boundaries and success criteria when requirements change, and it strengthens system contracts when implementation or review reveals a missing semantic constraint.
For changes that span multiple modules, dependency-aware specification patches identify the affected SpecModules and propagate updated guarantees through their dependents; unrelated modules retain their existing contracts.
The revised \spec{} returns to the implementation agent, closing the loop until the implementation satisfies the current blueprint.
Consequently, the same architecture governs both full-system construction and evolution in \sys{} and scoped development in Linux.

%% file: sections/design/design.tex
\section{Design}
\label{s:design}

\subsection{Bridging User Intent and OS Implementation with \spec{}}
\label{subs:development-blueprint}

\harness{} bridges high-level developer intent and low-level OS implementation through \spec{}, a persistent blueprint with two complementary components.
Starting from an often incomplete natural-language request, the specification agent uses iterative grilling and system analysis to construct \spec{} by co-developing a \emph{Plan} that bounds the current development task and a \emph{Specification} that captures the functionality, modularity, and concurrency semantics that its implementation must establish or preserve.
Because the initial intent does not fully determine either component, both are progressively refined as additional requirements and system constraints emerge.
This blueprint provides a persistent and inspectable basis for implementation, review, and subsequent refinement.

\myparagraph{Extending \oldspec{} to kernel-scale systems.}
SpecFS introduced \oldspec{}, a formal-method-inspired specification framework, and demonstrated that such specifications can guide LLMs in constructing and evolving a context-bounded user-space file-system subsystem.
\spec{} extends \oldspec{} by combining a task-bounding Plan with an OS-specific Specification.
Its Specification component retains the three complementary specification forms introduced by \oldspec{}---functionality, modularity, and concurrency specifications---but extends their scope to multi-module operating systems.
The Specification also supports incremental specification:
a task may describe only the semantic changes to an existing module, while unmentioned semantics remain inherited from its current specification.
Like \oldspec{}, \spec{} remains a lightweight and structured natural language representation for guiding implementation and review rather than a machine-checked proof language.

\myparagraph{Adding task boundaries through the Plan.}
The Specification component specifies the requested semantic change and the system contracts that its implementation must establish or preserve.
These contracts do not by themselves bound the development cycle, which also requires shared task objectives and exclusions, a permissible implementation scope, and observable completion criteria.
The Plan provides this task-level guidance without redefining the semantics captured by the Specification, preventing unnecessary changes and ensuring that the task does not terminate before its objectives are achieved.

\subsubsection{Functionality Specification}
\label{subs:design-function}

The functionality specification describes the state transitions of a module using pre- and post-conditions, system algorithms, and invariants.
The amount of algorithmic guidance scales with the complexity of the specified logic.
A \textbf{Level~1} specification uses pre- and post-conditions to define the allowed transition without prescribing its implementation.
A \textbf{Level~2} specification additionally provides a system algorithm when correctness, performance, or architectural constraints require a particular implementation strategy.
Invariants are orthogonal to these levels: they state properties that must remain true throughout execution whenever local entry and exit conditions alone are insufficient.

\myparagraph{Pre- and post-conditions.}
We adapt the Hoare-logic form $\{P\}C\{Q\}$ as a lightweight behavioral contract rather than a proof obligation requiring full formal verification.
The contracts are expressed in structured natural language, augmented with types and program expressions, so that they can directly guide implementation and review.
Pre-conditions define the states required before an operation may execute, while post-conditions guarantee its state transitions and return values.
For the user-space components targeted by \oldspec{}, assumptions over function arguments and module-local state were often sufficient.
Kernel behavior, however, can also depend on execution context, such as the current privilege level, interrupt or preemption state, address space, and global kernel state.
The functionality specification within \spec{} therefore extends these contracts to capture the relevant system state in addition to functional inputs.
For example, \autoref{fig:code:functionality-spec} specifies how \texttt{mem\_check} combines its socket argument with system-wide privilege and accounting state to decide whether another flow label may be allocated.

\input{sections/design/functionality-spec}

\myparagraph{System algorithm.}
Pre- and post-conditions constrain \emph{what} an operation must accomplish, but multiple implementations can satisfy the same state transition while having materially different performance or architectural consequences.
A system algorithm constrains \emph{how} the transition should be achieved by describing its major steps, data-structure choices, or required optimization strategy.
For example, the contracts for \texttt{kalloc()} and \texttt{kfree()} may only require allocating and returning a free page.
A single global free list satisfies that behavior but serializes allocation on a multicore system.
A Level~2 specification can instead require per-CPU free lists, with local allocation in the common case and cross-CPU page transfer only when a local list is exhausted, thereby preserving the same functional contract while constraining the implementation to a scalable design.

\myparagraph{Invariants.}
Invariants describe properties that must remain valid throughout an operation, rather than only at its entry and return.
They constrain the intermediate actions of every function that can affect the relevant state and therefore cannot be captured completely by local pre- and post-conditions.
For instance, an allocator must never place the same physical page on two free lists, and a page reachable through an active mapping must not simultaneously be considered free.
Stating these constraints separately avoids duplicating them across function contracts and makes them available to both implementation and review.

\subsubsection{Modularity Specification}
\label{subs:design-module}

The modularity specification defines how a module composes with the rest of the OS by making its external dependencies and exposed contract explicit.
At kernel scale, placing the entire codebase in an LLM's context is impractical, while reasoning about a module in isolation can miss assumptions imposed by other subsystems or privilege layers.
The modularity specification within \spec{} adapts rely-guarantee reasoning~\cite{10.1145/1480881.1480922} to make these composition boundaries explicit: each module states the external conditions on which its implementation depends and the contract that other modules may assume about it.
These contracts provide the context needed to implement a task-relevant module without loading or reconstructing the implementation details of the entire OS.

\myparagraph{Rely.}
A rely clause records assumptions supplied from outside the current module.
We distinguish two kinds of relies by the entity that discharges the assumption, rather than by the granularity at which the assumption is expressed.
A component rely is an assumption discharged by the guarantee of an identifiable SpecModule, whether it names a function, an interface, an entire module, or a semantic capability.
This generalizes \oldspec{}, in which component relies were typically expressed as function APIs with pre- and post-conditions, while preserving a concrete provider for each dependency.
A platform rely, by contrast, is a module-specific assumption about the selected execution substrate that the module consumes directly and that is not mediated by any SpecModule guarantee.
It therefore constrains the targets on which the module can execute rather than creating an edge in the SpecModule dependency graph.
For example, trap-entry code may rely directly on the architecture's trap semantics, whereas a module that consumes a trap representation supplied by an architecture module relies on that module's guarantee.

\myparagraph{Guarantee.}
A guarantee clause defines the externally visible contract that a module provides to its dependents, including its callable interfaces and their observable semantics.
This contract establishes an abstraction boundary: dependents may rely on the declared behavior, while internal functions, data structures, and implementation choices remain hidden.
Limiting dependencies to the guarantee prevents accidental coupling to kernel internals and allows a module's implementation to change as long as its exposed contract is preserved.
In \autoref{fig:code:modularity-spec}, for example, the IPv6 flow-label socket-option dispatcher relies directly on the GET, PUT, and RENEW handlers.
Its guarantee exposes a single dispatch interface that validates the request, selects the corresponding handler, and propagates its result.

\input{sections/design/modularity-spec}

\myparagraph{Controlled composition.}
For modules to compose, every component rely must be entailed by its identified provider's guarantee.
Every platform rely must instead hold on the selected execution target, thereby defining a compatibility constraint at the boundary of the SpecModule graph.
A kernel imposes stricter composition boundaries than a self-contained subsystem: modules cannot arbitrarily depend on internal behavior or expose contracts to every other component.
For example, a file system integrates with the VFS through registered operation interfaces such as \texttt{file\_operations} and \texttt{inode\_operations}, rather than relying on arbitrary VFS internals.
Across the privilege boundary, user programs can invoke kernel services only through the system-call ABI; internal kernel functions are not exposed as guarantees to user space.
Restricting composition to these explicit abstraction boundaries preserves privilege isolation, prevents unintended cross-module coupling, and allows internal implementations to evolve without forcing unrelated dependents to change.

\subsubsection{Concurrency Specification}
\label{subs:design-concurrency}

The concurrency specification defines how a module coordinates access to shared kernel state under concurrent execution.
Because kernel execution may interleave across CPUs, scheduler preemption, and interrupts, its synchronization constraints are difficult to infer from functional logic and costly to repeat for every function.
The concurrency specification within \spec{} therefore separates concurrency into locking protocols for reusable rules and locking logic for operation-specific synchronization.

\myparagraph{Locking protocol.}
A locking protocol defines the synchronization discipline shared by all operations that access the same state, independently of any one function's functional algorithm.
For modules with simple and uniform concurrency, the protocol alone can fully specify the required synchronization discipline.
In \autoref{fig:code:concurrency-spec}, the spinlock protocol requires every access to the global flow-label hash and its membership counters to hold \texttt{ip6\_fl\_lock}, with process and timer paths using the locking primitive appropriate to their execution context.
Stating this discipline once avoids repeating it in every operation that accesses the protected state.

\myparagraph{Locking logic.}
Locking logic specifies how a particular operation follows a protocol when the module-wide discipline alone does not determine the required critical-section boundaries.
In the same figure, the locking logic for \texttt{ipv6\_flowlabel\_get} places admission checking and insertion in one critical section, including a duplicate recheck before publishing a new label.
Separating this logic from allocation policy makes the synchronization behavior independently inspectable without repeating unrelated functional semantics.

\input{sections/design/concurrency-spec}

\subsubsection{Task-Bounding Plan}
\label{subs:design-plan}

Within \spec{}, the Specification defines the semantic change, whereas the task-bounding Plan governs the development cycle used to realize it.
It structures this task-level guidance into three components: goals and non-goals, scope, and success criteria.

\myparagraph{Goals and non-goals.}
Goals state the functional, performance, compatibility, or maintenance outcomes that the task must achieve.
Non-goals identify adjacent behavior that is outside the task and existing properties that must remain unchanged.
Making both explicit distinguishes a deliberate omission from a missed requirement and prevents an agent from introducing unrelated refactoring or optimizations while pursuing a locally reasonable implementation.

\myparagraph{Scope.}
Scope identifies the portions of the system that the task may change, such as subsystems, modules, files, symbols, or configuration, together with permitted new modules and dependencies and boundaries that must remain stable.
For a cross-module task, the scope can include components whose contracts must change to satisfy the task, as exposed by their rely-guarantee relationships.
Scope therefore limits implementation freedom without requiring the developer to specify every source-code edit in advance.

\myparagraph{Success criteria.}
Success criteria turn the goals into observable conditions under which the task is considered complete.
They may require a behavior or performance property, satisfaction of the relevant specification contracts, preservation of non-goals, and specific build, boot, test, or review outcomes.
The Plan records these required outcomes; \textsection\ref{subs:os-harness-agents} and \textsection\ref{subs:os-infrastructure} describe how \harness{} obtains the corresponding review and execution evidence.

The example in \autoref{fig:code:task-plan} bounds the IPv6 flow-label task to per-network-namespace admission and accounting, while preserving the existing flow-label interfaces and unrelated networking behavior.

\input{sections/design/task-plan}

\spec{} evolves as grilling reveals missing requirements and system analysis exposes implicit constraints.
Its Plan captures the boundary of the current task, while its Specification captures the task-relevant semantics that the implementation must establish or preserve.
Completing a task updates the affected components of \spec{} and the corresponding implementation, while specifications unrelated to the task remain unchanged.

\subsection{\harness{} Agents}
\label{subs:os-harness-agents}

As shown in \autoref{fig:design-overview}(b), \harness{} organizes OS development around four specialized agents: a specification agent, an implementation agent, a review agent, and a refinement agent.
These agents form three stages that progressively turn an incomplete request into an implementation and keep it aligned with evolving requirements and system constraints.
In the \emph{specification stage}, the specification agent grounds developer intent in the existing system and constructs \spec{} by producing its Plan and Specification.
In the \emph{implementation stage}, the implementation agent modifies code within the Plan's scope and verifies the result against the Specification and success criteria.
In the \emph{refinement stage}, the review agent independently examines the implementation, after which the refinement agent incorporates the review report and developer feedback into a revised blueprint for the next implementation iteration.
All four agents can draw on the shared OSKB, while the implementation agent invokes the OS Environment to obtain build and execution evidence; \textsection\ref{subs:os-infrastructure} describes these facilities in detail.

\subsubsection{Specification Agent}
\label{subs:specification-agent}

The specification agent helps the developer turn an incomplete natural-language intent into \spec{} by constructing a Plan and Specification that are sufficiently precise to guide implementation and review.
Because the initial request rarely contains every task boundary or system constraint, the agent constructs these components through an interactive analysis--grilling loop rather than a single translation step.

\myparagraph{Analysis--grilling loop.}
The agent first analyzes the intent together with the relevant code and existing artifacts.
This analysis locates affected components, recovers implicit contracts, and identifies missing or conflicting information that would make the intended change ambiguous.
The agent then grills the developer with focused questions about the unresolved requirements and design choices.
Each answer enriches the task context and triggers another round of code analysis, which may expose further dependencies or ambiguities.
The loop continues until the developer and agent reach a shared interpretation of the intended behavior, task boundary, and relevant system constraints.

\myparagraph{Artifact construction.}
From this progressively refined context, the agent authors \spec{} by producing its task-bounding Plan and corresponding Specification.
For an existing implementation, it can extract relevant behavior and contracts from code before introducing the requested changes; for a new component, it constructs the required contracts from the agreed design and surrounding system context.
Before handing the blueprint to the implementation agent, it checks its components for omissions, contradictions, and unresolved references between the Plan and affected specifications.

\subsubsection{Implementation Agent}
\label{subs:implementation-agent}

The implementation agent realizes the Specification as code within the constraints established by the Plan through an implement--verify loop.
This loop keeps code generation or modification within the task boundary and prevents a plausible implementation from advancing before it has been checked against the shared blueprint.

\myparagraph{Implementation.}
The Plan's scope determines which subsystems, modules, files, symbols, and dependencies the agent may change.
Within this boundary, the goals and non-goals constrain the intended effect of the modification, while the Specification governs its functional behavior, module composition, and concurrency semantics.
The agent uses these constraints to create a new implementation or revise the existing code without extending the change beyond the permitted scope.

\myparagraph{Verification.}
After each implementation attempt, the agent verifies that the code conforms to the relevant functionality, modularity, and concurrency specifications.
It then evaluates the Plan's success criteria using the required observable checks and evidence from the OS Environment.
A specification mismatch or unmet success criterion produces feedback for the next implementation attempt.
The implement--verify loop continues until the implementation conforms to the Specification and all applicable Plan success criteria are satisfied, after which the result is submitted to the review agent.

\subsubsection{Review Agent}
\label{subs:review-agent}

The review agent independently examines an implementation that has completed the implement--verify loop.
Separating review from implementation provides a fresh evaluation of both conventional code quality and OS-specific risks that may not be covered by the task's explicit checks.

\myparagraph{General review.}
General review identifies defects that do not require specialized OS knowledge, such as incorrect error handling, resource leaks, unsafe boundary conditions, and unnecessary changes outside the intended implementation logic.
It also checks whether the implementation remains understandable and consistent with the surrounding codebase.

\myparagraph{OS-specific review.}
OS-specific review grounds the analysis in domain knowledge retrieved from the OSKB.
We port Linux-derived review prompts into the OSKB to encode reusable review knowledge for kernel concerns such as synchronization, resource lifetime, user--kernel boundaries, and hardware interaction.
The review agent selects knowledge relevant to the affected subsystem and uses it to inspect risks that are easy to miss in general code review.
The construction and integration of these review prompts are described with the OSKB in \textsection\ref{subs:os-knowledge-base}.

\myparagraph{Review report.}
The agent produces a structured report that records each finding, its code location, the relevant OS rule or artifact, and a suggested refinement direction.
It does not directly modify the code or \spec{}; the report becomes an input to the refinement agent.

\subsubsection{Refinement Agent}
\label{subs:refinement-agent}

The refinement agent updates \spec{} by combining the review agent's report with developer feedback.
Its role is limited to refining this blueprint: it does not directly patch the implementation.

\myparagraph{Artifact refinement.}
Developer feedback may revise the task's goals, non-goals, scope, or success criteria, while review findings may reveal system semantics that should be clarified or strengthened in the Specification.
The agent incorporates both sources into an incremental update of the affected artifacts and leaves unrelated specifications unchanged.
The refined \spec{} is then returned to the implementation agent, whose implement--verify loop brings the code into alignment with the revised blueprint.

\subsection{OS Environment and Knowledge Base}
\label{subs:os-infrastructure}

\spec{} and specialized agents provide explicit development constraints, but they do not by themselves ground decisions in system execution or supply all of the domain knowledge needed for OS development.
\harness{} addresses these two gaps with shared infrastructure: the OS Environment validates implementations through actual build and execution, while the OS knowledge base (OSKB) supplies reusable, task-relevant knowledge to all four agents.

\subsubsection{OS Environment}
\label{subs:os-environment}

The OS Environment is an executable interface through which the implementation agent can build, boot, test, and debug a candidate OS implementation.
This execution boundary is essential because a module that appears correct in isolation may fail only when linked with the full kernel, exercised across user--kernel boundaries, or run with concurrent and hardware-facing components.
Textual reasoning and specification conformance are therefore complemented by validation of the complete executable system.

\myparagraph{Environment-backed end-to-end validation.}
The environment exposes a uniform set of operations for building the kernel, user runtime, and system image; booting the target OS in QEMU; running tests or workloads; and collecting console logs or GDB diagnostics when execution fails.
During the verification step, the implementation agent submits its current implementation to this environment, which builds and boots the complete target before executing the checks associated with the Plan's success criteria.
This process validates integration and runtime behavior rather than stopping at compilation or an isolated module test.
The OS Environment does not decide the intended behavior or prove conformance to the Specification; it executes observable checks and returns runtime evidence that the implementation agent combines with its specification-level analysis.

\myparagraph{Compact execution feedback.}
Build, test, and debugging workflows can produce output far larger than an agent's context window.
The OS Environment therefore returns a compact summary of the validation stage, pass/fail results, and relevant log or debugging evidence, while retaining complete raw output outside the agent context.
This summary provides enough information for the next implementation attempt without repeatedly consuming context with unrelated execution output.

\subsubsection{OS Knowledge Base}
\label{subs:os-knowledge-base}

The OSKB is a shared collection of reusable OS knowledge that supplies task-relevant context to all four agents.
Such knowledge is distributed across kernel code, subsystem documentation, review practices, and platform manuals, and cannot be placed in every agent context in full.
Centralizing it in the OSKB avoids requiring each agent to recover the same rules independently and helps maintain a consistent interpretation across the development workflow.

\myparagraph{Knowledge collection.}
We collect knowledge from kernel subsystem and API documentation, ISA and ABI specifications, hardware device and memory-model documentation, repository-specific coding and testing conventions, and project-maintained design constraints.
In particular, we port Linux-derived review prompts~\cite{linux-review-prompts} into the OSKB and organize them by subsystem and kernel concern, including synchronization, resource lifetime, user--kernel boundaries, and hardware interaction.
Collected knowledge retains its source and applicable subsystem, architecture, or concern so that agents can select rules appropriate to the current target.

\myparagraph{Agent integration.}
Agents retrieve OSKB knowledge using their role together with the relevant \spec{} context and the affected code.
Only the task-relevant subset is added to an agent's context.
The specification agent uses this knowledge to identify implicit constraints and formulate grilling questions; the implementation agent uses it to follow relevant APIs, platform rules, and repository conventions; the review agent uses Linux-derived prompts for OS-specific review; and the refinement agent uses the applicable rules to interpret findings and update \spec{}.

\subsection{Usage and Deployment of \harness{}}
\label{subs:usage-deployment}

\harness{} supports two complementary forms of use.
A developer can invoke any agent independently through a Rust command-line interface (CLI) and incorporate its output into an existing development workflow, or use an Orchestrator to execute the complete specification, implementation, review, and refinement process.

The same interfaces support the two development settings demonstrated in our case studies.
For \sys{}, \harness{} manages the artifacts and executable environment needed to construct and iteratively refine a complete OS.
For Linux kernel development, it operates on an existing repository under the corresponding task-scoped \spec{}, allowing the workflow to remain focused on the intended kernel change.

\myparagraph{Agent skills and Rust CLI.}
Each of the four agents is implemented as an independently callable skill that packages its role-specific instructions and tools.
We integrate these skills into a standalone Rust CLI that serves as the local runtime of \harness{}.
The CLI includes the OSKB and all agent skills, manages the kernel project's working directory, and persists the artifacts produced throughout a task, including \spec{}, verification and review reports, and other intermediate files.
A user can invoke an individual skill through the CLI, while every invocation operates on the same managed workspace and artifact state.

\myparagraph{Orchestrated workflow.}
The Orchestrator is a workflow controller rather than a fifth reasoning agent.
It uses the Rust CLI to invoke the specification, implementation, review, and refinement skills in sequence and passes the managed artifacts between them.
During specification, the Orchestrator pauses when the specification agent needs to grill the developer for missing information.
During refinement, it presents the review report, collects developer feedback, and invokes the refinement agent to update \spec{}.
Implementation, verification, and review can proceed automatically between these interaction points, and the refined \spec{} is returned to the implementation agent for the next iteration.

\myparagraph{Containerized deployment.}
We deploy \harness{} in a container that includes the Rust CLI, the Orchestrator, and the kernel development and validation toolchain.
Within the container, the CLI manages the OSKB, agent skills, kernel working directory, and task artifacts, while compilers, build tools, tests, QEMU, and GDB support executable validation.
The entire agent-driven kernel development workflow runs inside this environment.
Containerization therefore provides a reproducible and isolated execution boundary for both agent operation and kernel validation.

%% file: sections/design/functionality-spec.tex
\begin{figure}[H]
  \centering
  \captionsetup{aboveskip=15pt, belowskip=0pt}
  \begin{specblock}{FUNCTIONALITY SPECIFICATION}{text}
|\textbf{Pre-condition}|:
  |\textbf{Function arguments}|:
    - sk denotes a valid IPv6 socket.
  |\textbf{System state}|:
    - current's CAP_NET_ADMIN state determines whether unprivileged
      allocation limits apply.
    - fl_size, sk's flow-label list, and sock_net(sk)->flowlabel_count
      describe the current allocation state.

|\textbf{Post-condition}|:
|\textbf{Case 1}| Global pool exhausted
  - Return -ENOBUFS.
|\textbf{Case 2}| Unprivileged limit exceeded
  - Return -ENOBUFS if an applicable global-pressure, per-socket,
    or per-netns limit is reached.
|\textbf{Case 3}| Allocation admitted
  - Return 0.

|\textbf{Invariant}|:
  - fl_size equals global hash membership, and each flowlabel_count
    equals its network namespace's hash membership.
  \end{specblock}
  \caption{Simplified functionality specification for 
  \texttt{mem\_check}.}
  \vspace{-10pt}
  \label{fig:code:functionality-spec}
\end{figure}

%% file: sections/design/modularity-spec.tex
\begin{figure}[H]
  \centering
  \captionsetup{aboveskip=15pt, belowskip=0pt}
  \begin{specblock}{MODULARITY SPECIFICATION}{text}
|\textbf{[RELY]}|
ipv6_flowlabel_get(sk, freq, optval, optlen)
  // Handles GET and CREATE requests.
ipv6_flowlabel_put(sk, freq)
  // Releases the requested socket-owned flow-label reference.
ipv6_flowlabel_renew(sk, freq)
  // Renews the requested flow label's lifetime.

|\textbf{[GUARANTEE]}|
int ipv6_flowlabel_opt(struct sock *sk, sockptr_t optval, int optlen)
  // Copies and validates struct in6_flowlabel_req.
  // Dispatches GET, PUT, and RENEW to the corresponding handler.
  // Returns -EINVAL for an unsupported action; otherwise propagates
  // the selected handler's return value.
  \end{specblock}
  \caption{Rely--Guarantee specification for the IPv6 flow-label
  socket-option dispatcher.}
  \vspace{-10pt}
  \label{fig:code:modularity-spec}
\end{figure}

%% file: sections/design/concurrency-spec.tex
\begin{figure}[H]
  \centering
  \captionsetup{aboveskip=15pt, belowskip=0pt}
  \begin{specblock}{CONCURRENCY SPECIFICATION}{text}
|\textbf{[Spinlock Protocol: ip6\_fl\_lock]}|
|\textbf{Protected objects:}|
  - The global flow-label hash and its global/per-netns membership counters.
|\textbf{Rules:}|
  - All hash updates and counter accesses hold ip6_fl_lock.
  - Admission checking and insertion share one critical section.
  - Process paths use spin_lock_bh(); the GC timer uses spin_lock().

|\textbf{[Locking Logic: Serialized Admission and Insertion]}|
|\textbf{Pre-condition:}|
  - The candidate is unpublished and ip6_fl_lock is not held.
|\textbf{System algorithm:}|
  1. Enter an RCU read-side section and acquire ip6_fl_lock.
  2. Check admission and, if admitted, recheck for a duplicate;
     reuse it or publish the candidate and update both counters.
  3. Release ip6_fl_lock and leave the RCU read-side section.
|\textbf{Post-condition:}|
  - Admission and insertion are atomic with respect to other writers.
  - Hash membership and its counters remain consistent.
  - ip6_fl_lock and the RCU read-side section are not held.
  \end{specblock}
  \caption{Concurrency specification for serialized admission and insertion
  in \texttt{ipv6\_flowlabel\_get}.}
  \vspace{-10pt}
  \label{fig:code:concurrency-spec}
\end{figure}

%% file: sections/design/task-plan.tex
\begin{figure}[t]
  \centering
  \captionsetup{aboveskip=15pt, belowskip=0pt}
  \begin{specblock}{TASK-BOUNDING PLAN}{text}
|\textbf{[GOALS]}|
  - Prevent one unprivileged network namespace from exhausting the shared
    flow-label pool and starving other namespaces.
  - Keep admission and global/per-netns accounting exact under concurrency.

|\textbf{[NON-GOALS]}|
  - Do not change the flow-label UAPI, proc output, sharing modes,
    routing, or transport behavior.
  - Do not add a sysctl, netlink API, or hardware dependency.

|\textbf{[SCOPE]}|
  - net/ipv6/ip6_flowlabel.c: admission, insertion, and accounting.
  - include/net/netns/ipv6.h: per-netns flow-label count.

|\textbf{[SUCCESS CRITERIA]}|
  1. One unprivileged netns cannot exceed its per-netns budget.
  2. Insertion and removal update both counters exactly once; reuse does not.
  3. Admission and insertion execute in one critical section.
  4. Existing flow-label behavior and selftests remain compatible.
  \end{specblock}
  \caption{Simplified task-bounding Plan for enforcing an unprivileged
  per-network-namespace budget in the IPv6 flow-label manager.}
  \vspace{-10pt}
  \label{fig:code:task-plan}
\end{figure}

%% file: sections/evaluation/evaluation.tex
\section{Evaluation}
\label{s:eval}

We evaluate whether \harness{} can effectively support OS development in two complementary settings: constructing and evolving a new operating system, and implementing scoped changes in an existing production kernel.
The former requires \harness{} to maintain consistent semantics across a complete OS stack, whereas the latter requires it to recover and preserve the implicit constraints of a large, mature codebase.
Together, these settings allow us to evaluate both the system-wide and task-level effectiveness of \harness{}.

Our evaluation addresses the following research questions:
\begin{itemize}[leftmargin=*]
  \item \textbf{RQ1: Full-system construction.} Can \harness{} construct a complete, bootable operating system whose generated modules conform to their specifications and satisfy the system's functional requirements?
  \item \textbf{RQ2: Controlled system evolution.} Can \harness{} evolve an operating system through localized specification patches, including both cross-layer functionality and semantics-preserving performance optimizations?
  \item \textbf{RQ3: Existing-kernel development.} Can \harness{} complete representative Linux kernel development tasks more accurately than a general-purpose coding agent?
\end{itemize}

We answer these questions through two case studies.
First, we use \harness{} to construct \sys{}, a specification-guided reimplementation of xv6~\cite{cox2011xv6} consisting of 259 modules across the complete kernel and user-space stack.
We evaluate its construction through module-level generation accuracy, whole-system boot and functional tests, and comparisons with natural-language and oracle-context baselines across models of different capabilities.
We then evaluate its evolvability by extending the text-only system with a GUI stack and applying three performance optimizations through specification patches.
This case study answers RQ1 and RQ2 by examining whether \harness{} can both produce a complete OS and evolve it without rewriting unrelated parts of the system.

Second, we apply \harness{} to eleven representative development tasks drawn from the Linux kernel.
Unlike \sys{}, these tasks begin with an existing implementation whose relevant semantics, subsystem conventions, and compatibility constraints must be recovered from the codebase.
We evaluate whether the resulting patches correctly satisfy the task requirements and compare their development accuracy with those of a general-purpose coding agent.
This case study answers RQ3 and complements the full-system experiment by testing \harness{} in a mature kernel where changes must remain scoped and coexist with substantial existing functionality.

\input{sections/evaluation/specos-case-study.tex}
\input{sections/evaluation/linux-case-study.tex}

%% file: sections/evaluation/specos-case-study.tex
\subsection{Case Study: Constructing and Evolving \sys{}}
\label{subs:design-put-it-together}

\subsubsection{Building \sys{} with \harness{}}
\label{subs:building-specos}

\myparagraph{Overview.}
We use xv6-riscv as the basis for \sys{}, target its native RISC-V architecture, and run the generated system in QEMU.
Even at its modest scale, xv6 offers a sufficiently rich cross-subsystem setting for this case study.

\begin{figure*}[t]
  \centering
  \includegraphics[width=0.5\textwidth]{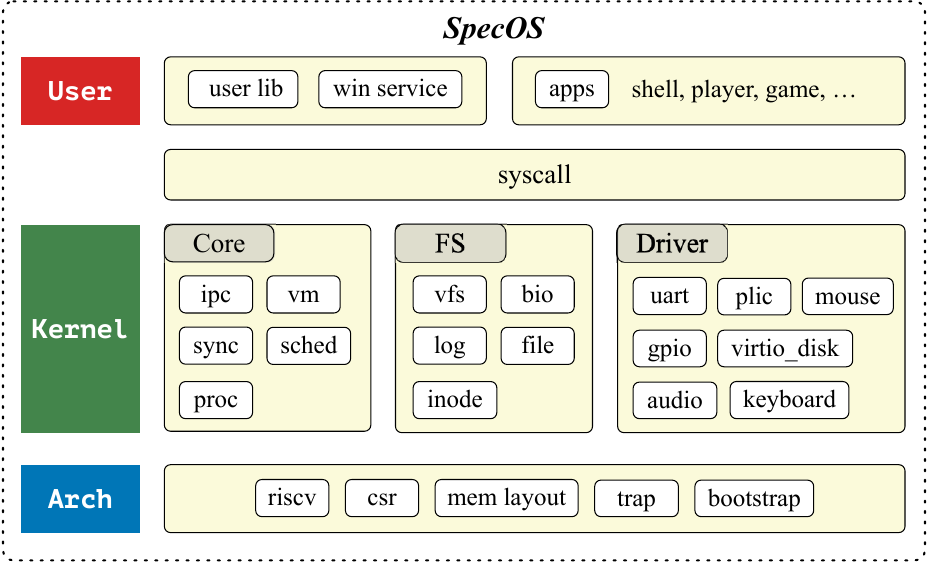}
  \setlength{\belowcaptionskip}{-5pt}
  \setlength{\abovecaptionskip}{5pt}
  \caption{\textbf{\sys{} architecture.}
  \sys{} comprises a kernel and a user-space stack organized into 11 logical components and 259 specification-governed modules.}
  \label{fig:specos-arch}
\end{figure*}

As shown in \autoref{fig:specos-arch}, \sys{} is organized into 11 logical components and 259 modules spanning memory management, process management, system calls, the file system, device drivers, kernel and user libraries, and low-level RISC-V routines.
The case-study artifact contains approximately 8K lines of \spec{} and 13K lines of generated C code.
We use C so that the generated implementation can be integrated directly with the xv6 build and compared with conventional low-level kernel code.

The construction stage receives the complete \sys{} \spec{} as a fixed input.
To isolate implementation from the other stages of \harness{}, we enable only the implementation agent and use its implement-verify loop to construct the system.

\myparagraph{Preparing the development blueprint.}
Before construction, the specification agent translates the original xv6 implementation into an initial full-system \spec{} comprising a Plan and Specification.
A developer participates in the grilling process to resolve underspecified behavior and design choices, and then inspects the resulting Specification to confirm that it captures the intended semantics.
The finalized \spec{} is fixed before implementation begins.

Rather than treating xv6 as a monolithic generation target, the Specification within the resulting \spec{} decomposes it into composable modules.
Each module records its pre- and post-conditions, invariants, rely/guarantee contracts, algorithmic description, and concurrency requirements.
These specifications expose the contracts needed to implement a module while preserving its relationships with the rest of the system.

\myparagraph{Generating the implementation.}
The implementation agent consumes the complete \sys{} \spec{} and applies an \emph{implement-verify} loop.
In each iteration, it produces a candidate C implementation and verifies the candidate against the corresponding specification.
A failed verification returns feedback to the next implementation attempt; a module advances only after satisfying the verification step.
Applying this loop across all 259 modules produces the implementation that we integrate and evaluate as \sys{}.

\subsubsection{Functional Correctness and Generation Accuracy}
\label{s:eval-accuracy}

\myparagraph{End-to-end functional correctness.}
We first evaluate whether the generated modules compose into a working operating system.
The resulting \sys{} boots on QEMU, runs the standard xv6 user-space environment, and passes all 70 registered test cases in the functional test suite of xv6.
These tests exercise the integrated kernel through its user-visible interfaces and show that \harness{} can generate a complete, executable system. 
They provide end-to-end functional evidence rather than a proof of correctness beyond the tested behaviors.

\myparagraph{Module-level accuracy.}
We separately evaluate whether each generated module preserves the semantics of its corresponding ground-truth implementation.
The evaluation covers all 259 modules in \sys{}.
We use GPT-5.4 as an LLM judge to compare each generated module with its ground truth.
A module is counted as correct only when the judge determines that the two implementations are semantically and logically equivalent; any omitted behavior, inconsistent state transition, or incorrect implementation logic is counted as a failure.
We report accuracy as the fraction of the 259 modules judged correct.

We compare the Specification component of \spec{} with two baselines:
\begin{itemize}[leftmargin=*]
  \item \textbf{Normal.} The prompt provides a natural-language description of the module logic together with the signatures and type definitions of the dependency APIs it may require.
  \item \textbf{Oracle.} In addition to the Normal input, the prompt provides the complete ground-truth source code of the dependency modules.
  \item \textbf{Specification from \spec{} (our approach).} The prompt uses the structured module specification, including behavioral contracts, state transitions, and inter-module dependencies.
\end{itemize}
We run all three methods with Qwen3.6-Plus, GPT-5.4, and Qwen3.6-35B-A3B to cover models with different capabilities.

\begin{figure}[t]
  \setlength{\belowcaptionskip}{-10pt}
  \setlength{\abovecaptionskip}{0pt}
  \centering
  \includegraphics[width=0.4\textwidth]{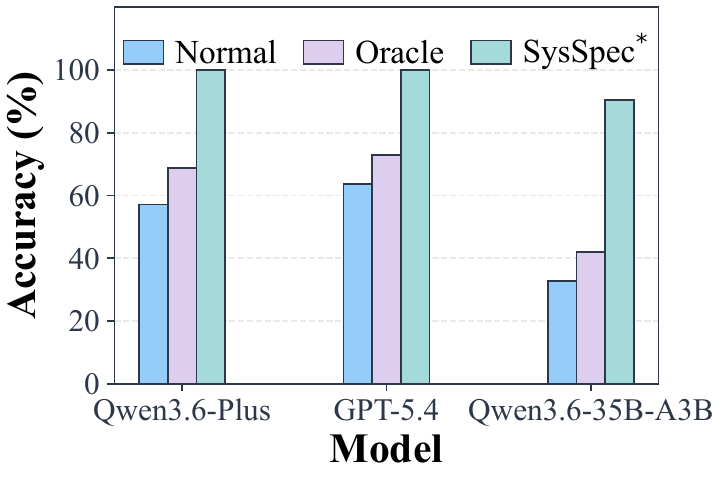}
  \caption{\textbf{Module-level generation accuracy for \sys{}.}}
  \label{fig:eval-accuracy}
\end{figure}

\myparagraph{Accuracy results.}
As shown in \autoref{fig:eval-accuracy}, the Specification consistently improves module-level accuracy across all three models.
With Qwen3.6-Plus and GPT-5.4, it achieves 100\% accuracy: all 259 generated modules are judged semantically equivalent to their ground-truth implementations.
Both the Normal and Oracle baselines remain less reliable, indicating that natural-language requirements and additional dependency source code do not provide the same effective constraints as the structured specification.

The improvement remains substantial with the smaller Qwen3.6-35B-A3B model, for which the Specification achieves 90.3\% accuracy and outperforms both baselines.
It also exceeds the Oracle results obtained with stronger models.
These results suggest that explicit behavioral and inter-module constraints can compensate in part for lower model capability.
Together with the successful boot and functional tests, the module-level results provide complementary evidence that \harness{} can generate both individually conforming modules and a complete working operating system.

\subsubsection{Evolving \sys{} with \harness{}}
\label{sec:case-study-gui}

\myparagraph{Overview.}
We next evaluate whether \harness{} can evolve a working OS from a user-level intent.
Starting from the request to turn the text-only \sys{} into a GUI-capable system, the developer and the specification agent jointly produce the final \spec{}, including its Plan and relevant Specification content, through an iterative analysis--grilling loop that combines code analysis with focused questions about requirements and design choices.
\harness{} then generates and verifies the implementation and incorporates review findings through refinement.
This task crosses device drivers, memory management, system calls, user-space services, libraries, and applications.
We organize the evolution into three consecutive phases: kernel support, a user-space window service, and application adoption.
We validate the result by running graphical applications and the existing xv6 functional tests.

\myparagraph{Phase I: Kernel support.}
The original \sys{} provides only a text console.
The first phase adds a virtio-gpu driver with a scanout-capable frame buffer and a graphics ABI for querying display geometry and submitting full-screen or partial updates.
To prevent applications from competing for the display, the kernel grants presentation authority to the window service; the kernel controls the device, while the service determines composition policy.
We also add shared-memory IPC so that clients can exchange frame buffers with the service without repeatedly transferring pixel data through byte-oriented pipes.
The shared-memory subsystem tracks mappings per process and cleans them up on \texttt{exec()} and \texttt{exit()}.
Finally, event-oriented keyboard input allows the kernel to capture key events for subsequent delivery to the focused application.

\myparagraph{Phase II: Window service.}
The second phase introduces a window protocol, a user-space compositor, and a client library.
The protocol defines commands, events, and shared-memory buffers exchanged between clients and the service.
Each window uses double buffering: a client renders into a back buffer and commits it, after which the compositor presents the new buffer and releases the previous one.
This avoids concurrent modification of a buffer being composed.
Started by \texttt{init}, the compositor maintains window state and z-order, presents the composed frame through the graphics ABI, and routes input to the focused client.
The client library exposes this service through a simple window API, hiding the underlying protocol and shared-memory details.

\myparagraph{Phase III: Application adoption.}
The final phase adds an SDL compatibility layer over the client library.
The SDL compatibility layer maps SDL surface updates and input events to the window protocol, allowing existing SDL applications to be ported to \sys{} with only minor source-level changes.
To validate GUI functionality from complementary perspectives, we select a representative set of applications with different interaction patterns. 
Some of these applications are adapted from the Navy-apps suite~\cite{navyapps}.
Collectively, they cover the end-to-end rendering path from application buffers through shared-memory transport and composition to GPU display, together with the input path from the kernel to the focused client.
The game repeatedly renders frames and responds to input; the launcher creates windows and starts applications; the multi-window workload presents concurrent windows under changing focus; and the graphical terminal accepts keyboard input, renders text, and executes commands.

\begin{figure}[t]
  \centering
  \begin{subfigure}[t]{0.125\textwidth}
    \centering
    \includegraphics[height=0.13\textheight,keepaspectratio]{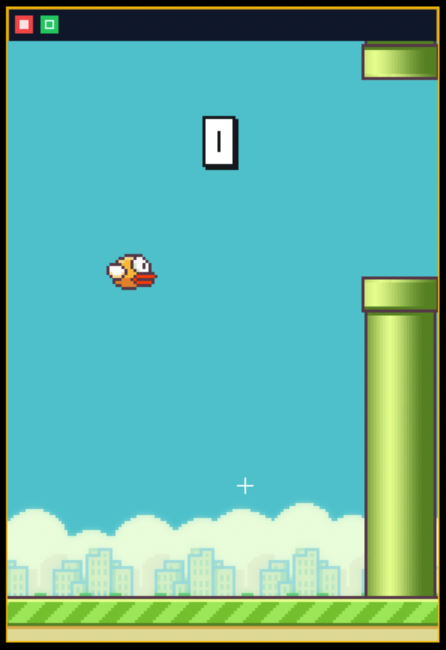}
    \caption{Game}
  \end{subfigure}
  \hfill
  \begin{subfigure}[t]{0.29\textwidth}
    \centering
    \includegraphics[height=0.13\textheight,keepaspectratio]{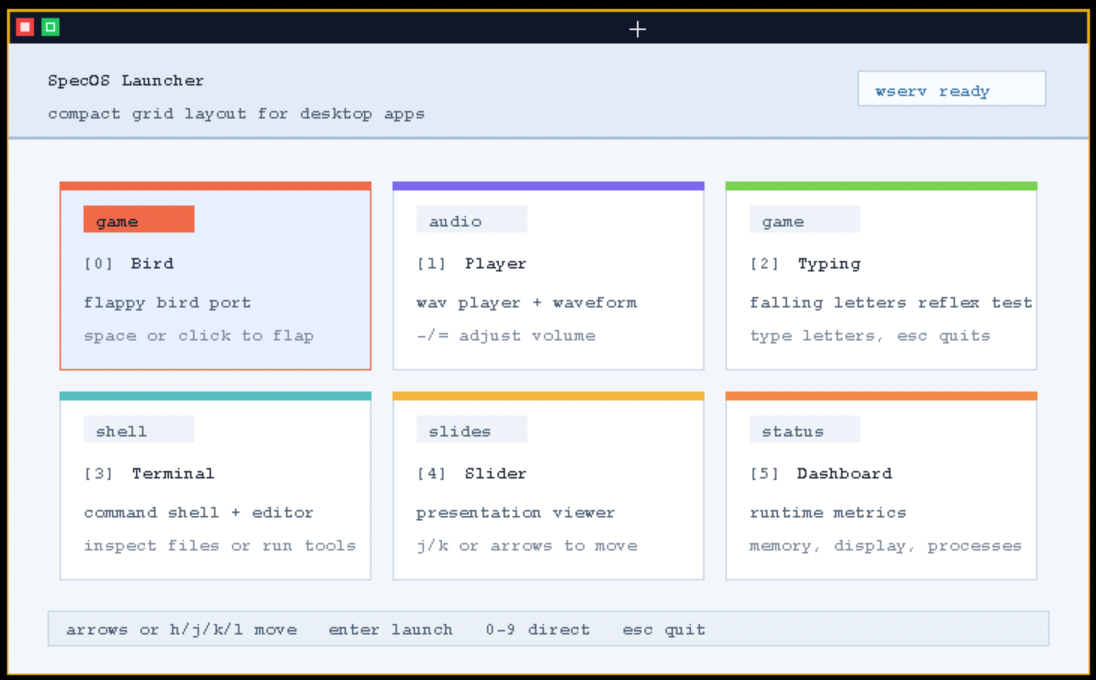}
    \caption{Launcher}
  \end{subfigure}
  \hfill
  \begin{subfigure}[t]{0.265\textwidth}
    \centering
    \includegraphics[height=0.13\textheight,keepaspectratio]{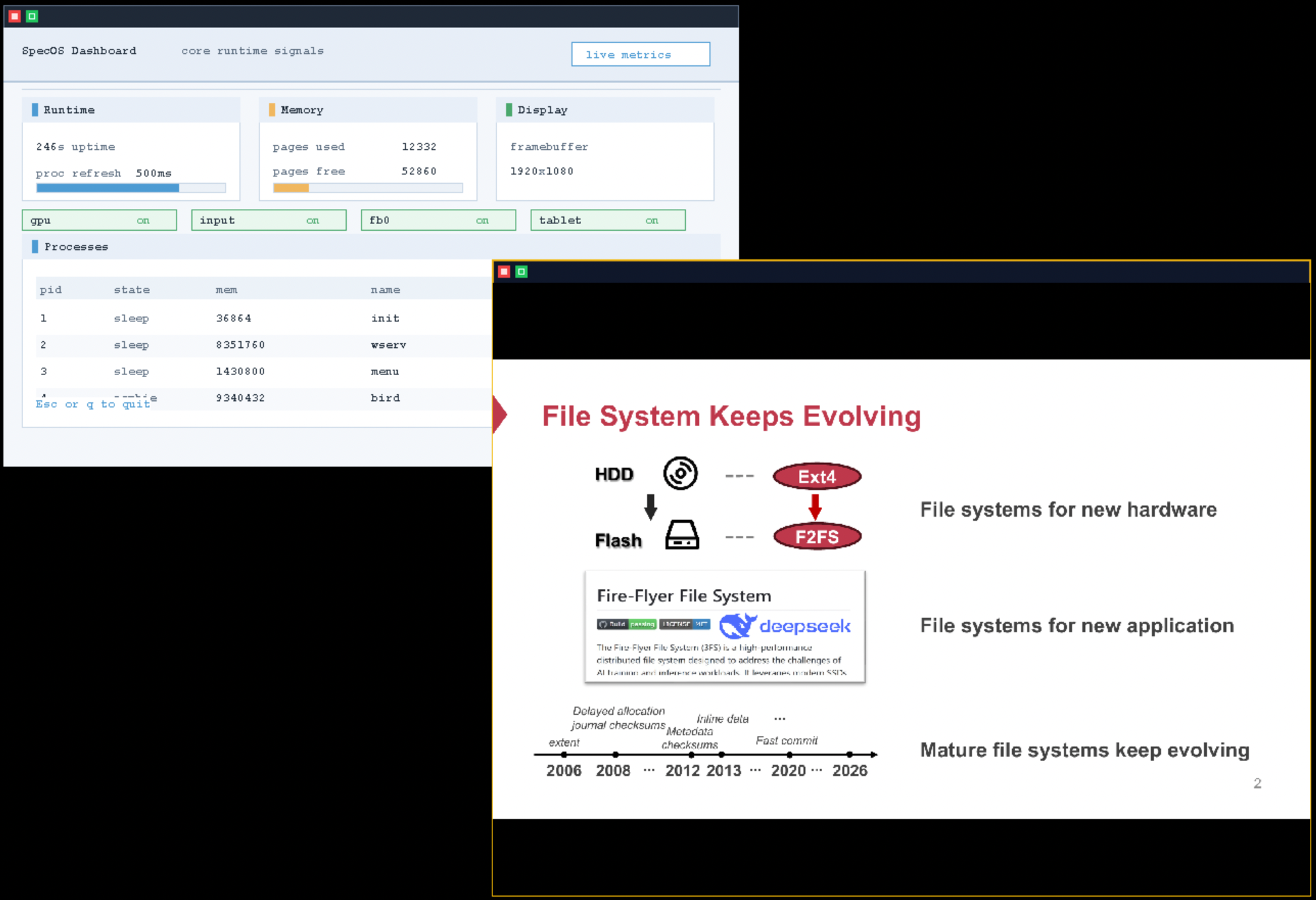}
    \caption{Multi-window}
  \end{subfigure}
  \hfill
  \begin{subfigure}[t]{0.295\textwidth}
    \centering
    \includegraphics[height=0.13\textheight,keepaspectratio]{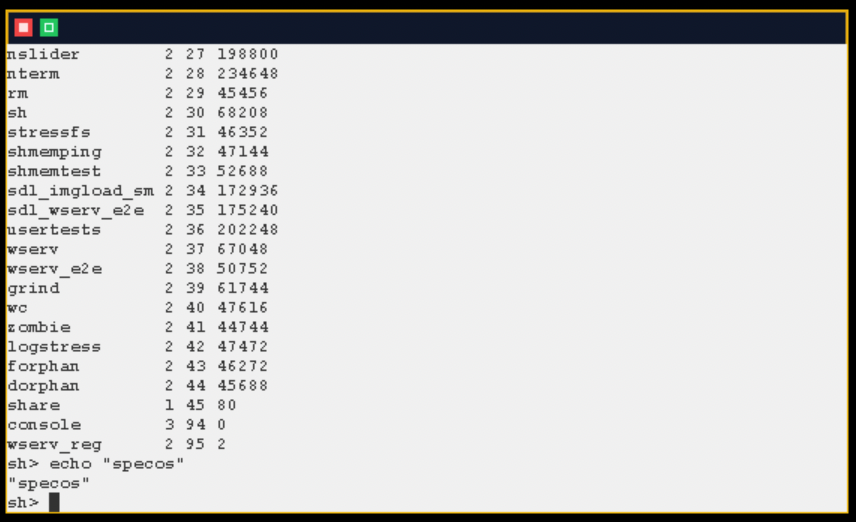}
    \caption{Terminal}
  \end{subfigure}
  \setlength{\belowcaptionskip}{-5pt}
  \setlength{\abovecaptionskip}{5pt}
  \caption{\textbf{End-to-end GUI applications running on the evolved \sys{}.}
  \textit{The screenshots show a game, the window-service launcher, multiple concurrent windows, and a graphical terminal.}}
  \label{fig:case-study-gui-apps}
\end{figure}

\myparagraph{Validation.}
The evolved system boots into the GUI environment and runs the game, launcher, multi-window workload, and graphical terminal shown in \autoref{fig:case-study-gui-apps}.
Their execution validates display output, shared-memory transport, composition, and input delivery across the new stack.
The system also continues to pass the complete xv6 functional test suite, providing regression evidence that the evolution preserves the existing behaviors covered by those tests.

\subsubsection{Performance Evaluation}
\label{subs:specos-performance}

We further evaluate whether \harness{} can translate performance intents into effective system optimizations.
We provide separate intents for a per-CPU page allocator, copy-on-write (COW) fork, and batched pipe I/O.
For each intent, \harness{} updates the Plan and relevant parts of the Specification within \spec{}, generates and verifies the affected implementation, and incorporates review feedback.
We compare each generated variant with the original generated \sys{} under the same QEMU environment.
The optimizations change internal data structures or algorithms while retaining their user-visible interfaces, and all three variants continue to pass the xv6 functional test suite.

\begin{figure}[t]
  \setlength{\belowcaptionskip}{-10pt}
  \setlength{\abovecaptionskip}{0pt}
  \centering
  \begin{minipage}[t]{0.3\linewidth}
    \centering
    \includegraphics[width=\textwidth]{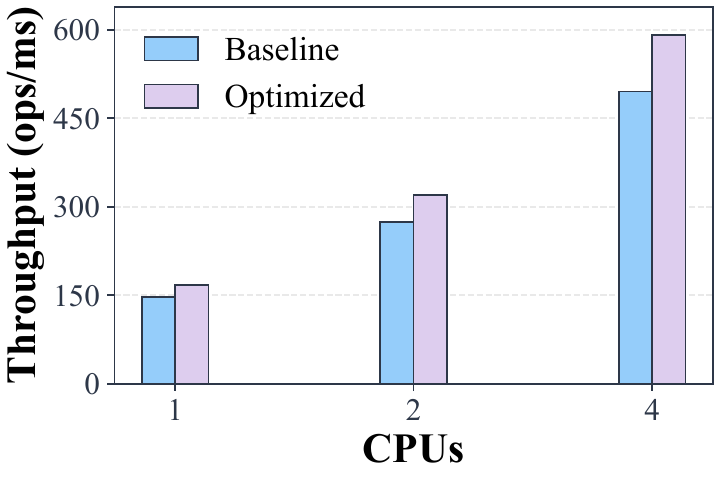}
  \end{minipage}
  \hfill
  \begin{minipage}[t]{0.3\linewidth}
    \centering
    \includegraphics[width=\textwidth]{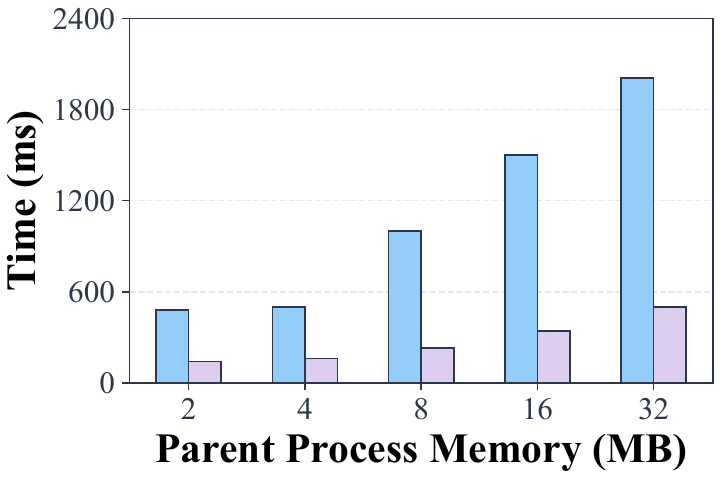}
  \end{minipage}
  \hfill
  \begin{minipage}[t]{0.3\linewidth}
    \centering
    \includegraphics[width=\textwidth]{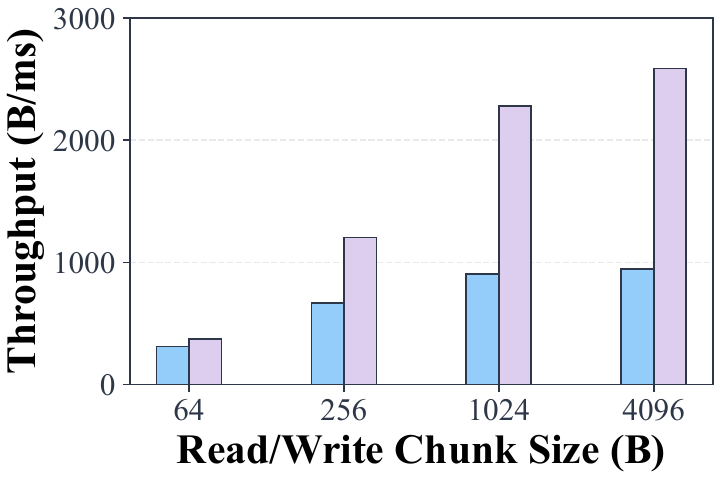}
  \end{minipage}
  \caption{\textbf{Performance improvements from \harness{}-generated \sys{} optimizations.}
  \textit{Left: the per-CPU page allocator reduces global allocator contention as concurrency increases.
  Middle: copy-on-write fork avoids eager page copying in fork-exec workloads.
  Right: batched pipe I/O reduces pipe transfer overhead, especially for larger block sizes.}}
  \label{fig:eval-os-optimizations}
\end{figure}

\myparagraph{Per-CPU page allocator.}
The baseline allocator maintains one free list protected by a global spinlock, so every \texttt{kalloc} and \texttt{kfree} operation accesses shared state.
The generated optimization gives each CPU a local free list and serves common allocations and frees from that list.
Local lists exchange pages with a global reserve only when they require a refill or drain, and batched allocation/free interfaces amortize the remaining allocator operations.
Our benchmark repeatedly allocates and frees user memory under 1, 2, and 4 CPUs.
The benefit grows with concurrency, reaching a 19.2\% improvement at 4 CPUs; under 2 and 4 CPUs, the measured spin count on the global \texttt{kmem} lock falls to zero per 1,000 operations.
Because the optimization introduces both local lists and batching, the result reflects their combined effect and is consistent with moving the common allocation path away from the global lock.

\myparagraph{Copy-on-write fork.}
The baseline \texttt{fork()} allocates and copies every user page before the child begins execution.
The generated COW implementation instead lets the parent and child share physical pages, marks their mappings as copy-on-write, and maintains per-page reference counts.
A private copy is allocated only when either process writes to a shared page, preserving the independent-address-space semantics of \texttt{fork()} while deferring copying.
Our fork-exec workload allocates and touches a parent address space before repeatedly forking children that immediately execute a new program.
This workload exposes pages that eager fork copies but the child discards at \texttt{exec}.
Across address-space sizes from 2 MB to 32 MB, COW avoids these copies and achieves a 3.12$\times$ to 4.41$\times$ speedup over the baseline.

\myparagraph{Batched pipe I/O.}
The baseline pipe path invokes \texttt{copyin} or \texttt{copyout} separately for each byte, making large transfers pay repeated user--kernel copy overhead.
The generated implementation computes a contiguous chunk from the requested length, available pipe space or data, and ring-buffer boundary, and transfers that chunk with one copy operation.
It retains the pipe's synchronization and user-visible read/write semantics while changing the granularity of data movement.
We measure transfer time between a writer and reader while varying the user-level block size from 64 to 4,096 bytes.
Batching improves performance at every tested size, with larger blocks benefiting more from the reduced number of copy calls.
At 4096 bytes, it reaches a 2.74$\times$ speedup and reduces transfer time by 63.5\%.

Together, these experiments show that \harness{} can realize performance intents that require changes to data structures and internal algorithms, while the regression results preserve the existing behaviors covered by the xv6 test suite.

%% file: sections/evaluation/linux-case-study.tex
\subsection{Case Study: Linux Kernel Development}
\label{subs:linux-case-study}

\subsubsection{Evaluation Setup}
\label{subs:linux-eval-setup}

\myparagraph{Tasks.}
We select eleven bug-fix tasks from Linux kernel commits merged during the three months preceding our evaluation.
All selected commits postdate the release of the evaluated model, reducing the likelihood that their fixes appeared in its training data.
The tasks span networking, file systems, memory management, security, and tracing, addressing systems concerns ranging from synchronization to resource isolation and validation.
The task suite ranges from single-file fixes to substantial changes spanning multiple files.
\autoref{tab:linux-tasks} summarizes the task suite and outcomes.

\begin{table*}[t]
  \centering
  \small
  \caption{\textbf{Linux kernel bug-fix task suite and outcomes.}}
  \label{tab:linux-tasks}
  \begin{tabularx}{\textwidth}{@{}Xllccc@{}}
    \toprule
    \textbf{Task} & \textbf{Subsystem} & \textbf{Core issue} & \textbf{Files} & \textbf{Codex} & \textbf{\harness{}} \\
    \midrule
    IPv6 flow-label limits          & Networking        & Lock synchronization  & 2 & \textcolor{green}{\ding{51}} & \textcolor{green}{\ding{51}} \\
    ext4 fast-commit wakeup         & Filesystems       & Wakeup liveness       & 2 & \textcolor{green}{\ding{51}} & \textcolor{green}{\ding{51}} \\
    hugetlb poison deadlock         & Memory management & Deadlock avoidance    & 4 & \textcolor{green}{\ding{51}} & \textcolor{green}{\ding{51}} \\
    SELinux audit cache             & Security          & Audit consistency     & 2 & \textcolor{green}{\ding{51}} & \textcolor{green}{\ding{51}} \\
    ESP shared fragments            & Networking        & Fragment marking      & 4 & \textcolor{green}{\ding{51}} & \textcolor{green}{\ding{51}} \\
    eprobe error offset             & Tracing           & Diagnostic location   & 1 & \textcolor{red}{\ding{55}} & \textcolor{green}{\ding{51}} \\
    request-key concurrent revoke   & Security          & Object lifetime       & 4 & \textcolor{red}{\ding{55}} & \textcolor{green}{\ding{51}} \\
    lockd NLM file leak             & Filesystems       & Resource cleanup      & 3 & \textcolor{red}{\ding{55}} & \textcolor{green}{\ding{51}} \\
    fanotify permission events      & Filesystems       & Permission isolation  & 3 & \textcolor{red}{\ding{55}} & \textcolor{green}{\ding{51}} \\
    IPVS resize livelock            & Networking        & Concurrent liveness   & 2 & \textcolor{red}{\ding{55}} & \textcolor{red}{\ding{55}} \\
    NTFS validation                 & Filesystems       & Metadata validation   & 4 & \textcolor{red}{\ding{55}} & \textcolor{red}{\ding{55}} \\
    \bottomrule
  \end{tabularx}
\end{table*}

\myparagraph{Compared approaches.}
For each task, both conditions start from the same base commit and receive the same \texttt{task.md}, which specifies the goal, required behavior, and correctness criteria.
Both run in the isolated \emph{os-env} described in \textsection\ref{subs:os-environment} and use GPT-5.5, configured with high reasoning effort, as the LLM backbone.
We use the general-purpose Codex coding agent as the baseline.
The baseline operates directly on the task, whereas, for this evaluation, \harness{} generates \spec{} directly from \texttt{task.md} in a single pass, without the analysis--grilling loop, and then coordinates implementation, verification, review, and refinement.
No human feedback is introduced during patch generation; instead, the original \texttt{task.md} is supplied as a fixed feedback source during refinement alongside the review report.
Thus, the development harness is the primary difference between the two conditions.

\myparagraph{Success criterion.}
We validate a candidate with its task-specific test when one is available; otherwise, an LLM judge compares the candidate and ground-truth patches under the behavioral requirements in \texttt{task.md}.
The judge rejects omitted behavior, semantic deviations, and changes outside the required behavior.
We count a task as successful only when its corresponding validator accepts the candidate.

\subsubsection{Lessons from Representative Cases}
\label{subs:linux-case-lessons}

Three representative tasks in which the baseline fails but \harness{} succeeds illustrate complementary benefits of explicit task constraints, independent review, and OS-specific knowledge.

\myparagraph{\spec{} prevents implementation drift.}
The eprobe task requires the parser to preserve the error offset already computed for a missing attached-event field; the existing error class and all other parser behavior must remain unchanged.
The baseline locates the offset reset but also assigns a new error code, turning a location fix into an unsupported semantic change.
In contrast, within the \harness{} \spec{}, the Plan records preservation of the error class as a non-goal, while the Specification requires the existing failure path to retain the computed offset.
By following these \spec{} constraints, the implementation agent checks that the neighboring parser branches remain unchanged and produces a patch that removes only the offset reset.
The judge therefore accepts the \harness{} patch and rejects the baseline patch for changing behavior outside the task.

\myparagraph{Independent review provides actionable guidance.}
The request-key task must keep an authorization payload alive while instantiate and reject operations may sleep concurrently with revoke or destroy.
The baseline implements the core reference-counting design but misses a key condition on the repeated-revoke path, failing to account for the case in which the authorization payload has already been cleared.
The initial \harness{} implementation also misses this repeated-revoke condition.
The review agent then independently performs a path-complete check of the authorization-record lifecycle: revoke must detach the payload before releasing its owner reference, and every instantiate/reject exit path must release the pinned reference exactly once.
This review exposes the lifecycle condition missed by the initial \harness{} implementation.
The refined patch correctly implements these conditions on all relevant paths and passes the judge.
Although independent review does not guarantee correctness, it provides a separate opportunity to detect overlooked constraints and turn them into concrete refinement guidance.

\myparagraph{OSKB captures OS-specific semantics and concerns.}
The fanotify task requires permission-event preparation to preserve mark lifetime and group isolation while the \texttt{fsnotify\_mark\_srcu} read-side protection is temporarily released for a user-space wait.
The baseline incorrectly aborts the current permission event on a stale mark from an unrelated group instead of skipping it under SRCU.
OSKB provides task-relevant rules for SRCU-protected mark traversal, reference lifetime, and concurrency semantics.
Guided by these rules, \harness{} skips such marks, still fails when a current-group mark cannot be retained, and balances references across the wait.
The resulting patch correctly handles the relevant paths and passes the judge.

\subsubsection{Failure Analysis}
\label{subs:linux-failure-analysis}

We next analyze the two tasks that neither the baseline nor \harness{} completes to identify current boundaries of \harness{}.

\myparagraph{IPVS resize livelock.}
The IPVS task requires table resizing to complete while updates and readers continue to access the service table.
The ground-truth design uses \texttt{svc\_replace\_sem} to coordinate table replacement, ensuring consistent publication and RCU-safe reclamation of the old table.
\harness{} recognizes the local risk that a blocking table walker may repeatedly restart.
However, it does not infer that resizing and concurrent updates must be coordinated as one table-replacement operation.
Its \spec{} therefore permits a design based on \texttt{svc\_resize\_sem} and \texttt{service\_mutex}.
These locks serialize selected paths, but they do not ensure that the new table is published consistently or that the old table remains alive until readers finish.
The Plan states the correct liveness goal, but the Specification fails to specify the synchronization protocol required to achieve it.
This case illustrates a limitation of \harness{}: when a task leaves important semantics implicit, especially in its concurrency logic, \harness{} may fail to derive a correct \spec{}, leaving subsequent agents unable to complete the task correctly.

\myparagraph{NTFS metadata validation.}
The NTFS task requires rejecting attribute-list entries whose \texttt{name\_offset} does not immediately follow the fixed entry header.
Thus, \texttt{name\_offset} must equal \texttt{sizeof(struct attr\_list\_entry)}, rather than merely fall within the record.
\harness{} derives the record-boundary check but does not recover the stricter layout invariant relating the on-disk offset to the fixed header.
The resulting implementation therefore prevents out-of-bounds access but accepts offsets that remain within the record while pointing to the wrong location, so the parser may read the attribute name from an invalid position.
This case illustrates a limitation of \harness{}: it currently cannot reliably infer domain invariants that are implicit and distributed across data representations and their consumers.

%% file: sections/related-work.tex
\section{Related Work}
\label{s:related-work}

\myparagraph{General software engineering and coding agents.}
Advances in LLMs have led to a growing number of general-purpose coding agents, such as Codex~\cite{codex}, Claude Code~\cite{claudecode}, Pi-agent~\cite{piagent}, and Kimi Code~\cite{kimicode}, that assist software engineering by navigating repositories, editing source code, invoking build and test tools, and iterating on the resulting feedback.

Alongside the development of these general-purpose agents, substantial research has explored how to improve their capabilities and use them more effectively in software engineering.
AutoCodeRover~\cite{zhang2024autocoderover}, CodeAgent~\cite{zhang2024codeagentenhancingcodegeneration}, and RepairAgent~\cite{bouzenia2025repairagent} use structured repository search, tool integration, and iterative validation for issue resolution, repository-level generation, and program repair.
Agentless~\cite{xia2024agentless}, by contrast, shows that a fixed localization--repair--validation pipeline can remain competitive without autonomous tool use or action planning.

Benchmarking has evolved in parallel from isolated coding tasks toward long-horizon software engineering, ranging from resolving real-world repository issues to constructing complete systems from scratch.
Benchmarks such as SWE-bench Pro~\cite{deng2026swebench}, ProgramBench~\cite{yang2026programbench}, and DeepSWE~\cite{huang2026deepswe} evaluate agents across this broader spectrum of complex, multi-step development tasks.
Together, these systems and benchmarks demonstrate that coding agents are moving beyond local code completion toward sustained software engineering, although their semantic context is still primarily derived from natural-language task descriptions and existing repositories.

Another line of work investigates specification-driven workflows~\cite{piskala2026spec} for general-purpose software development.
Spec Kit~\cite{githubspeckit} and Kiro~\cite{kiro} maintain requirements, specifications, plans, and implementation tasks as persistent artifacts throughout the development process.
Because these approaches are designed for general software engineering, they do not prescribe how domain-specific OS semantics should be represented and maintained across development stages.
\harness{} specializes this workflow for systems software by making \spec{}, with its Plan and OS-specific Specification, a shared artifact across specification, implementation, review, and refinement.

\myparagraph{LLM-assisted systems software development.}
Recent work has begun to characterize the capabilities required for LLM-assisted systems engineering.
SEC-bench Pro~\cite{lee2026sec} evaluates long-horizon vulnerability reproduction in browser engines and the Linux kernel, while the system-intelligence proposal~\cite{feng2025systemintelligence} emphasizes reasoning about architectures, protocols, abstractions, and design tradeoffs beyond code generation and repair.
These efforts define and evaluate systems capabilities rather than provide an end-to-end development workflow.

Most existing systems-oriented approaches apply LLMs to a specific development activity.
For reliability, WASABI~\cite{stoica2024wasabi} combines LLM-based analysis with static and dynamic techniques to detect retry bugs, while KernelGPT~\cite{yang2025kernelgpt} synthesizes and iteratively validates syscall specifications for kernel fuzzing.
For Linux-kernel debugging, kAgent~\cite{mathai2026kagent} uses crash logs and reproduction feedback to generate and refine patches, Code Researcher~\cite{singh2025code} retrieves relevant context from source code and commit history, and LinuxFL+~\cite{zhou2026taming} improves fault localization.

Performance-oriented work is similarly specialized.
SysGPT~\cite{park2025principles} provides optimization suggestions grounded in a taxonomy of serial optimization strategies, while gigiprofiler~\cite{318449} combines LLM-based semantic inference with static and runtime analysis to diagnose application-defined resource bottlenecks.

Other performance-oriented systems directly optimize implementations.
ECO~\cite{318445} automatically localizes, generates, and validates efficiency patches in production code, while CUDA Agent~\cite{dai2026cuda} uses agentic reinforcement learning and execution feedback to optimize GPU kernels.
QiMeng-GEMM~\cite{zhou2025qimeng} uses architecture-aware prompt search to generate GEMM implementations, while QiMeng-Xpiler~\cite{qimengxpiler} combines LLM transformations, symbolic repair, and auto-tuning for cross-platform tensor programs.
Together, these approaches support performance analysis and feedback-driven code optimization, but target individual patches, kernels, operators, or transformations rather than the construction and evolution of a complete system.

Work at a broader scope explores the agentic construction of complete systems software.
SpecFS~\cite{315951} uses structured specifications to guide the construction and evolution of a concurrent file system, while SpecDB~\cite{lou2026specdb} generates customized database systems through feature-oriented decomposition.
VibeServe~\cite{kamahori2026vibeserve} uses nested agent loops to generate LLM serving stacks specialized to a target model, hardware platform, and workload.
\harness{} extends SpecFS to full-OS construction and scoped kernel evolution, while keeping specifications synchronized with code through implementation, execution, and review feedback.

\myparagraph{Formal methods and LLM-assisted verification.}
Formal methods have long been used to construct and verify complex systems.
seL4~\cite{klein2009sel4}, CertiKOS~\cite{gu2016certikos}, and Hyperkernel~\cite{nelson2017hyperkernel} demonstrate machine-checked verification of operating-system kernels.
Related work verifies file systems such as FSCQ~\cite{chen2016fscq}, AtomFS~\cite{zou2019atomfs}, and RefFS~\cite{zou2024reffs}, as well as liveness properties of cloud controllers such as Anvil~\cite{sun2024anvil}.
These systems provide substantially stronger guarantees than testing or informal review, but require carefully designed formal models, specifications, and proofs that are expensive to construct and maintain.

Recent work uses LLMs to reduce the effort required to construct formal specifications and proofs.
OSVBench~\cite{li2025osvbenchbenchmarkingllmsspecification} evaluates the generation of complete state-transition specifications for Hyperkernel, while SpecGen~\cite{ma2025specgenautomatedgenerationformal} generates and verifier-checks formal specifications for Java programs.
AutoVerus~\cite{yang2025autoverus} and SAFE~\cite{chen2025automated} generate correctness proofs for Rust through verifier-guided refinement or self-evolution, whereas AlphaVerus~\cite{aggarwal2025alphaverus}, Refine4LLM~\cite{10.1145/3704905}, and Clover~\cite{sun2024cloverclosedloopverifiablecode} integrate formal verification into code generation and refinement.
These approaches produce specifications, proofs, or programs that are checked against formal verification conditions.

By contrast, \harness{} does not generate machine-checked proofs or claim formal correctness.
Instead, it borrows the formal-methods discipline of making system behavior and cross-module contracts explicit, while using compilation, execution, and specification-aware review for lightweight validation.

%% file: sections/conclusion.tex
\section{Conclusion}
\label{s:conclusion}

This paper presented \harness{}, an OS-specialized agent harness that coordinates specification, implementation, review, and refinement around \spec{}, a shared development blueprint comprising a task-bounding Plan and an OS-specific Specification.
For full-system construction and evolution, \harness{} generated a complete, bootable \sys{} that passes its functional test suite, and subsequently evolved it through specification patches to support a GUI and multiple performance optimizations.
For existing-kernel development, it achieved 1.8$\times$ Codex's accuracy on 11 Linux-kernel bug-fix tasks spanning multiple Linux-kernel subsystems.
Together, these results show that explicit and persistent OS semantics can support both full-system generation and scoped development of existing kernels.